\documentclass[11pt]{article}

\usepackage{amssymb}

\usepackage{graphicx}

\usepackage[english,francais]{babel}

\newfont{\ensmathquatorze}{msbm10 scaled 1400}
\newfont{\ensmathonze}{msbm10 scaled 1100}
\newfont{\ensmathdix}{msbm10}
\newfont{\ensmathneuf}{msbm10 scaled 833}
\newfont{\ensmathhuit}{msbm10 scaled 694}
\newfam\ensmathfam                        
\textfont\ensmathfam=\ensmathonze        
\scriptfont\ensmathfam=\ensmathdix       
\scriptscriptfont\ensmathfam=\ensmathhuit
\def\ensmf{\fam\ensmathfam\ensmathonze}         

\def\be{\begin{equation}}
\def\ee{\end{equation}}

\def\bea{\begin{eqnarray}}
\def\eea{\end{eqnarray}}
\def\beann{\begin{eqnarray*}}
\def\eeann{\end{eqnarray*}}

\def\typ{\mbox{\tiny typ}}

\renewcommand{\geq}{\geqslant}

\newcommand{\ket}[1]{|\kern.3ex#1\kern.3ex\rangle}
\newcommand{\bra}[1]{\langle\kern.3ex #1 \kern.3ex|}
\newcommand{\APPROX}[1]{                
   {{\raisebox{-.3cm}{$\textstyle\simeq$}} \atop {\scriptstyle{#1}}}}

\newcommand{\mean}[1]{\left\langle #1 \right\rangle} 
\newcommand{\smean}[1]{\langle #1 \rangle} 

\newcommand{\EXP}[1]{{\mbox{\large e}}^{#1}}         
\renewcommand{\cosh}[1]{\mathop{\mathrm{ch}}\nolimits #1} 
\renewcommand{\sinh}[1]{\mathop{\mathrm{sh}}\nolimits #1} 

\newcommand{\re}{\mathop{\mathrm{Re}}\nolimits}      
\newcommand{\im}{\mathop{\mathrm{Im}}\nolimits}      
\newcommand{\tr}[1]{\mathop{\mathrm{Tr}}\nolimits\left\{ #1 \right\}}  
\newcommand{\cotg}{\mathop{\mathrm{cotg}}\nolimits}  

\def\NN{{\ensmf N}}                 
\def\ZZ{{\ensmf Z}}                 
\def\RR{{\ensmf R}}                 

\def\I{{\rm i}}                  
\def\D{{\rm d}}                  
\def\Dc{{\rm D}}                 

\newcommand{\dz}{\partial_{z}}

\newcommand{\drond}[2]{\frac{\partial #1}{\partial #2}} 

\newcommand\dphi{{\cal D}\phi{\cal D}\bar\phi\:}
\newcommand\lab{l_{\alpha\beta}}
\newcommand\lamab{\lambda_{\alpha\beta}}
\newcommand\lamba{\lambda_{\beta\alpha}}
\newcommand\Lamab{\Lambda_{\alpha\beta}}
\newcommand\Lamba{\Lambda_{\beta\alpha}}
\newcommand\Oab{{\cal O_{\alpha\beta}}}

\newcommand{\diagram}[3]{\raisebox{#3}{\includegraphics[scale=#2]{#1}}}

\begin{document}

\selectlanguage{english}

\title{Spectral determinant on quantum graphs}

\author{Eric Akkermans$^{\dagger,\star,\ddagger}$, Alain Comtet$^\ddagger$,
        Jean Desbois$^\ddagger$, \\
	Gilles Montambaux$^\star$ and Christophe Texier$^{\ddagger,\#}$}

\date{\today}

\maketitle	

{\small
\noindent
$^\dagger$
Department of Physics. Technion, Israel Institute of Technology, 
32000 Haifa, Israel.

\noindent
$^\star$
Laboratoire de Physique des Solides. 
Universit\'e Paris-Sud, B\^at.~510, F-91405 Orsay Cedex, France.

\noindent
$^\ddagger$
Laboratoire de Physique Th\'eorique et Mod\`eles Statistiques.
Universit\'e Paris-Sud, B\^at. 100, F-91405 Orsay Cedex, France.

\noindent
$^\#$
D\'epartement de Physique Th\'eorique, Universit\'e de Gen\`eve, 24~quai 
Ernest Ansermet, CH-1211 Gen\`eve~4, Switzerland.
}

\begin{abstract}
We study the spectral determinant of the Laplacian on finite graphs 
characterized by their number of vertices $V$ and of bonds $B$. We present 
a path integral derivation which leads to two equivalent expressions of the 
spectral determinant of the Laplacian either in terms of a $V\times V$ vertex 
matrix or a $2B\times2B$ link matrix that couples the arcs (oriented bonds) 
together. This latter expression allows us to rewrite the spectral determinant 
as an infinite product of contributions of periodic orbits on the graph. 
We also present a diagrammatic method that permits us to write the spectral 
determinant in terms of a finite number of periodic orbit contributions.
These results are generalized to the case of graphs in a magnetic field. 
Several examples illustrating this formalism are presented and its 
application to the thermodynamic and transport properties of weakly 
disordered and coherent mesoscopic networks is discussed. 
\end{abstract}


\section{Introduction and main results\label{Intro}}

This work is devoted to the study of the spectral properties of the Laplacian 
operator on finite graphs. This problem has already a long history. The 
properties of the Laplacian describing free electrons on networks made of 
one-dimensional wires have been investigated in the context of organic 
molecules \cite{RudSch53}. 
Subsequently, this approach has proved useful to study superconducting 
networks using linearized Ginzburg-Landau equations \cite{Ale83}, the 
vibration properties of fractal structures such as the Sierpinski gasket 
\cite{Ram84}, the adiabatic conductances of a network in an inhomogeneous
magnetic field \cite{AvrSad91,Avr95} or the behaviour of disordered systems 
in a magnetic field \cite{DouRam85,DouRam86,Mon96b,PasMon98,Pas98,PasMon99}. 
More recently, it has been shown \cite{KotSmi97,KotSmi99,KotSmi99a} that the 
Laplacian on graphs provides an interesting framework to study the onset of 
quantum manifestations of chaos. In these examples, the quantity of main 
interest is the spectrum of eigenenergies of the Schr\"odinger equation, or 
the diffusion equation, defined for each wire of the network with appropriate 
boundary conditions at the vertices. 
Finally we should also mention the relevance of the 
spectral determinant of the Laplacian on graphs in the context of statistical 
field theory and more precisely for the problem of the triangulation of random 
surfaces. There, the corresponding homology and (dual) cohomology groups 
provide the corresponding graphs over which the Laplacian is defined 
\cite{ItzDro89a}.

The problem of the spectrum of graphs has been also investigated thoroughly in 
the mathematical literature \cite{Big76,CveDooSac80,Chu97,Col98,Car99}. 
The more 
specific question of deriving a trace formula for the heat kernel (partition
function) of the Laplacian operator on a graph has been studied by Roth 
\cite{Rot83,Rot83a}.  

\bigskip

Let us start by recalling some already known results.
The partition function $Z(t)$ of the scalar Laplacian 
operator $\Delta$ with appropriate boundary conditions, is defined 
as
\be
Z(t) = \tr{\EXP{t \Delta}}
\:.\ee
Evaluating the trace over a set of eigenstates of $-\Delta$ of eigenvalues 
$E_n$, we obtain
\be
Z(t)=\sum_n \EXP{-E_n t}
\:.\ee
The spectral determinant $S(\gamma)$ is formally defined by 
$S(\gamma)= \det(-\Delta+\gamma) = \prod_n(\gamma+E_n)$ up to a 
regularization independent of $\gamma$. It is such that 
\be\label{relZS}
\int_0^\infty\D{t}\, Z(t) \EXP{- \gamma t}=\frac{\partial}{\partial\gamma} 
\ln S(\gamma)
\:.\ee
On a graph in the presence of a magnetic field, a compact form for the 
determinant of the operator $\gamma - (\D_x - \I A(x))^2$  has recently been 
obtained \cite{PasMon99}:
\begin{equation}
S(\gamma)  = \gamma^{V-B \over 2} \
\prod _{(\alpha \beta)} \sinh(\sqrt{\gamma }l_{\alpha\beta}) \:\det(M)
\:,\label{detpasmon} 
\end{equation}
where
$V$ is the number of vertices, $B$ is the number of bonds and $\lab$ is the 
length of the bond $(\alpha\beta)$. $M$ is a $V \times V$ matrix whose 
elements are \cite{Ale83,DouRam85,DouRam86,PasMon99}:
\bea\label{MatrixM1}
M_{\alpha\alpha}&=&\sum_{i=1}^{m_\alpha}\coth(\sqrt{\gamma}l_{\alpha\beta_i}) 
\\
\label{MatrixM2}
M_{\alpha\beta} &=&
               -\frac{\EXP{\I\theta_{\alpha\beta}}}{\sinh(\sqrt{\gamma}\lab)}
\ \ \mbox{ if $(\alpha\beta)$ is a bond} \\
                &=& 0 \ \ \mbox{ otherwise}\nonumber
\:,\eea
where the summation is taken over the $m_\alpha$ neighbouring sites of 
the vertex $\alpha$. 
$\theta_{\beta\alpha} = \int_\alpha ^\beta A(x)\D{x}$ is a phase 
equal  to the circulation of the vector potential $A(x)$ along the 
bonds $(\alpha\beta)$. This expression has been obtained by a calculation 
of the Green's function on the graph. The derivation is recalled in appendix 
\ref{DemGilles}.

Although it is well known that the spectrum of the Laplacian on a graph is 
given by the zeros of $\det(M(\gamma=-E))=0$ \cite{RudSch53,Ale83,Avr95},
formula (\ref{detpasmon}) is richer since it contains non trivial 
multiplicative factors which depend on $\gamma$. If the relation of 
$S(\gamma)$ to the matrix $M(\gamma)$  allows a very simple determination of 
the spectrum, it cannot help  to understand the structure of the spectrum in 
terms of periodic orbits on the graph. To do so, it is more convenient to 
introduce a "link matrix" $E$ whose elements $E_{(\alpha\beta)(\mu\nu)}$ 
relate oriented bonds $(\alpha\beta)$ and $(\mu\nu)$. This 
$2B \times 2B$ matrix is defined as\footnote{
The matrix $E$ is the same as the matrix $S_B$ in \cite{KotSmi99}}:
\bea
E_{(\alpha\beta)(\mu\nu)}
  &=& \frac{2}{m_\beta}
      \EXP{-\sqrt{\gamma}\,l_{\alpha\beta}+\I\theta_{\alpha\beta}}
      \ \ \mbox{ if } \beta=\mu \label{MatrixE1}\\
  &=& \left(\frac{2}{m_\beta}-1\right)
      \EXP{-\sqrt{\gamma}\,l_{\alpha\beta}+\I\theta_{\alpha\beta}} 
      \ \ \mbox{ if } \left\{\begin{array}{l}
                             \beta=\mu\\ \alpha=\nu
                             \end{array}\right. \label{MatrixE2}\\
  &=& 0 \ \ \mbox{ otherwise}    
\:.\eea
The spectrum is now given by $\det(1-E(\gamma))=0$. 
The trace expansion of $\det(1-E)$ leads to a  description of the spectrum 
in terms of periodic orbits \cite{KotSmi99}.

Finally, a trace formula has been obtained by Roth, which expresses the 
partition function as a contribution of such periodic orbits:
\be\label{RothIntro}
Z(t)=\frac{L}{2\sqrt{\pi t}}+\frac{V-B}{2}+\frac{1}{2\sqrt{\pi t}}
\sum_{C} l(\tilde C)\alpha(C)\EXP{-\frac{l(C)^2}{4t}}
\:,\ee
where the sum runs over closed orbits $C$ on the graph. $\tilde C$ is 
the primitive orbit associated with $C$. The geometrical factor 
$\alpha(\tilde C)$ will be defined in section \ref{TraFor}

\bigskip

We now give an overview of the paper and describe rapidly our main results. 
In section \ref{Meso} we recall briefly how transport and thermodynamic 
properties of weakly disordered and coherent conductors can be related to
the spectral determinant \cite{Mon96b,PasMon98,Pas98}. In particular the 
weak-localization correction to the conductance is a physical measure of the 
spectral determinant.

After having set up the notations and definitions in section \ref{DefNot},
we present in section \ref{PathInt} a path-integral formulation for the 
calculation of the spectral determinant of the Laplacian on a graph. It 
allows us to obtain an expression of the spectral determinant in terms of the 
vertex matrix (\ref{MatrixM1},\ref{MatrixM2}). The interest of this approach 
relies on the fact that the path-integral on each bond involves the propagator 
of a two-dimensional harmonic oscillator for which the role of the time is 
played by the length along the bond. This problem corresponds to a 
zero-dimensional Gaussian field theory, so that the path-integral is easily 
calculated using standard quantum mechanics. This approach has many 
advantages like to allow a simple generalization to other types of boundary 
conditions, or to permit very easily the elimination of vertices of 
coordinence $2$ (appendix \ref{loops}).

In section \ref{Jean}, we derive a dual and equivalent expression of the 
spectral determinant in terms of the link matrix 
(\ref{MatrixE1},\ref{MatrixE2}):
\be
S(\gamma)=\gamma^{\frac{V-B}{2}}\EXP{\sqrt{\gamma}\,L}
\Big(\prod_{\alpha}m_\alpha\Big)2^{-B}\ \det(1-E(\gamma))
\:.\ee
By a trace expansion of the determinant $\det(1-E)$, we find in section 
\ref{TraFor} the following result
\be\label{SIntro}
S(\gamma)=\gamma^{\frac{V-B}{2}}\EXP{\sqrt{\gamma}\,L}
\Big(\prod_{\alpha}m_\alpha\Big)2^{-B}\ 
\prod_{\tilde C}\left(
  1-\alpha(\tilde C)\EXP{-\sqrt{\gamma}\,l(\tilde C)+\I\theta(\tilde C)}\right)
\:,\ee
which allows us to recover  the Roth trace formula for the the partition 
function (\ref{RothIntro}), demonstrating the equivalence between 
(\ref{detpasmon}) and (\ref{RothIntro}).
The trace formula (\ref{SIntro}) for the spectral determinant 
involves the contribution of an infinite number of periodic orbits 
$\tilde{C}$ and bears some similarity 
with the Selberg trace formula \cite{Sel56} or the semi-classical trace 
formulae for chaotic Hamiltonian systems \cite{Gut90}. 
In section \ref{DiagExp} we show that the spectral determinant involves
the contributions of a finite number of periodic orbits only. This provides a 
diagrammatic method to compute the determinant. This method is applied to
a particular example.

In section \ref{Bfield} the path integral formulation is shown to be a good 
starting point to generalize these results to the case where a 
magnetic field is applied to the network. In this way we obtain a 
generalization of the Roth formula (\ref{RothIntro}).
Sections \ref{MixBoun} and \ref{ZeroMode} discuss how to extend the previous 
results to the case of more general boundary conditions, and provide a 
discussion of the low energy behaviour of the spectrum and the existence of 
zero modes, respectively.

Section \ref{Scatt} discusses the scattering problem when the graph is 
connected to an infinite lead. In this case the relevant information is 
contained in a phase shift which is shown to be related to the ratio of two 
spectral determinants calculated with different boundary conditions. 
Finally, section \ref{CompGra} gives an illustration of the formalism
for the case of the complete graph $K_n$.


\section{Phase coherent properties of disordered conductors \label{Meso}}

In this section, we briefly recall how transport and thermodynamic properties 
of weakly disordered and coherent conductors can be related to the partition 
function $Z(t)$ or equivalently to the spectral determinant $S(\gamma)$. 
A more detailed derivation can be found in references 
\cite{Mon96b,PasMon98,Pas98} and references therein.
These expressions are quite general and are not specific to graphs. But their 
calculation on any network made of quasi-one-dimensional diffusive wires is 
straightforward with the expression (\ref{detpasmon}) \cite{PasMon99}. 

\medskip

The first step is to write the physical quantities in terms of the classical 
return probability $P(\vec r,\vec r,t)$ solution of the diffusion equation 
(we set $\hbar=c=1$)
\be\label{diff2}
\left[ {\partial \over \partial t} - 
      D\left({\vec\nabla} + 2 \I e {\vec A}\right)^2
\right] P(\vec r,\vec r\, ',t)=  \delta(\vec r-\vec r\,')  
\ee
where $-e$ is the electron charge, $D$ is the diffusion coefficient. 
In the other sections of this paper, we have set $-2e=1$ and $D=1$. 
The return probability has actually two contributions, a purely classical 
$P_{cl}$ which is solution of Eq. (\ref{diff2}) with ${\vec A}=0$ and another 
one $P_{int}$ which  is the result of constructive interferences between 
electronic propagators associated to time-reversed trajectories. This second 
contribution only exists when the system is phase coherent. In the presence 
of a magnetic field, time reversed trajectories accumulate opposite phases. 
This is why $P_{int}$ obeys Eq. (\ref{diff2}) with an effective charge 
$2e$ \cite{Ber84}. In this section, we use a unique notation for the two 
contributions.

The solution of the diffusion equation (\ref{diff2}) has the form
\be
P(\vec r,\vec r\,',t)=\theta_{\rm H}(t) 
\sum_n \psi_n(\vec r) \psi_n^*(\vec r\,') \EXP{- E_n t}
\ee
where $\theta_{\rm H}(t)$ is the Heaviside function. 
The eigenvalues $E_n$ and the eigenfunctions $\psi_n$ are solutions of 
\be\label{schrodinger}
-D \left({\vec\nabla} + 2 \I e {\vec A} \right)^2
   \psi_n(\vec r) = E_n \psi_n(\vec r)
\:.\ee
The integral over the whole space of the return probability is precisely the 
partition function of the Laplace operator $-\Delta$, or 
$-({\vec \nabla} + 2\I e {\vec A})^2$ in a magnetic field, namely
\be
Z(t)=\int\D\vec r\, P(\vec r,\vec r,t) = \sum_n e^{- E_n t}
\:.\ee
As recalled in the introduction (Eq. (\ref{relZS})), the Laplace transform 
of $Z(t)$ is related to the spectral determinant $S(\gamma)$. 

\medskip

The electrical conductivity is a current-current correlation function and 
expresses in 
terms of products of two electron propagators. These propagators have 
uncorrelated phase factors which cancel after disorder averaging. This 
corresponds to the classical Drude-Boltzmann
conductivity $\sigma_0=  e^2 D \rho_0$, where  $\rho_0$ is the density of 
states at Fermi energy. However, pairs of time reversed trajectories have the 
same action and thus the same phase. 
This constructive interference leads to an additional 
contribution to the conductivity, called the weak-localization correction
\cite{Ber84} which is proportional to the probability of having pairs of time 
reversed trajectories, {\it i.e.} to the probability for a diffusive particle 
to return to the origin. The correction can be written as
\be\label{WL} 
\mean{\Delta\sigma}
= \langle \sigma \rangle - \sigma_0= - {2   e^2  \over \pi } 
{D \over \Omega}\int_0^\infty\D{t}\, Z(t) \EXP{-\gamma t} 
\ee
with $\Omega$ being the volume of the system. The exponential damping at large 
time is due to the breakdown of phase coherence because of inelastic 
processes. This breakdown is phenomenologically described by a characteristic 
time $\tau_\phi=1/\gamma$, related to the phase coherence length\footnote{
Integrals (\ref{WL},\ref{UCF},\ref{currentee3},\ref{Mtyp}) may not converge 
at small time. In this case the lower bound of the integral should be the 
elastic  time $\tau_e$, the smallest time for diffusion.} 
$L_\phi=\sqrt{D \tau_\phi}$. A magnetic field  or a Aharonov-Bohm flux,  
by breaking the time-reversal symmetry, destroys the weak-localization 
correction.

\medskip

In a mesoscopic system, {\it i.e.} with a typical size $L$ smaller than 
$L_\phi$, 
the conductivity is known to exhibit large variations from sample to sample, 
with a variance which is universal in the limit $L/L_\phi \rightarrow 0$. The 
structure of this variance results from different phase correlations between 
four electron propagators (two for each conductance). It can be shown that 
this variance can also be written in the following compact form 
\cite{PasMon98}:
\be
\smean{\delta\sigma^2}
=  {12  e^4   \over \beta \pi^2 }
{D^2 \over \Omega^2} \int_0^\infty \D{t}\,t \ Z(t) 
\EXP{-\gamma t}
\label{UCF} 
\ee
where $\beta=1$ if there is time reversal symmetry  and $\beta=2$ in the 
absence of such symmetry. 

\medskip

The magnetization due to the orbital motion of the non-interacting electron 
gas is known as the Landau magnetization. Taking into account 
electron-electron interactions in the Hartree-Fock picture gives an 
additional contribution, known as the Aslamasov-Larkin correction 
\cite{AslLar74}. The structure of this  correction implies the product of 
four wave functions, {\it i.e.} two local density of states whose disorder  
averaged product can be related to the Fourier transform of the return 
probability $P({\vec r},{\vec r},t)$ \cite{Mon96}. After spatial integration, 
one gets:
\be\label{currentee3} 
\langle M_{ee} \rangle =  -{\lambda_0
 \over  \pi} {\partial \over \partial {\cal B}} 
\int_0^\infty\D{t}\,  Z(t,{\cal B}){ \EXP{-\gamma t} \over
t^2} 
\ee
where $\lambda_0$ is the dimensionless interaction parameter\footnote{
Considering higher corrections in the Cooper channel leads 
to a ladder summation \cite{AslLar74,Eck91}, so that $\lambda_0$ should be 
replaced by $\lambda(t)  = \lambda_0/(1+\lambda_0
\ln(\epsilon_F t))=1/\ln(T_0 t)$ where $T_0$ is defined as $T_0=
\epsilon_F \EXP{1/\lambda_0}$. The authors of reference 
\cite{UllRicBarOppJal97} found $\lambda(t)=1/\ln(\epsilon_F t)$.
}. $\vec{\cal B}={\vec \nabla} \times {\vec A}$ is the magnetic field.

\medskip

Finally the typical
magnetization $M_{\typ}$, defined as $M_{\typ}^2 =
 \langle M^2 \rangle - \langle M \rangle^2$, can also be written in terms of  
field derivatives of $Z(t)$. Neglecting electron-electron interactions, the 
magnetization is proportional to the density of states. Thus the typical 
magnetization involves the two-point correlation function of the density of 
states which has been shown by Argaman {\it et al} \cite{ArgImrSmi93} to be 
related to the Fourier transform of $t Z(t)$. 
One obtains:
\be\label{Mtyp}
M^2_{\typ}  = \frac{1}{2\pi^2}
\int_{0} ^{\infty}\D{t}\, [Z''(t,{\cal B})-Z''(t,0)] {e^{-\gamma t} \over t^3}
\ee
where $Z''(t,{\cal B})=\partial^2 Z(t,{\cal B}) / \partial {\cal B}^2$.

\medskip

Using standard properties of Laplace transforms, the above time integrals can 
be written as integrals of the logarithm of the spectral determinant, so that 
the  physical quantities described above can be written as \cite{PasMon99}:
\bea
\mean{\Delta\sigma} 
  &=&   - {2e^2 \over \pi} {D \over \Omega}
\ \ \ \ {\partial \over \partial \gamma} \ln S(\gamma)        \\
\smean{\delta\sigma^2}
  &=& -{12e^4 \over \beta\pi^2}\frac{D^2}{\Omega^2}\ \  \drond{^2}{\gamma^2} 
      \ln S(\gamma)                                           \\
M^2_{\typ}  
  &=& \frac{1}{2\pi^2}\ \ \ \int_\gamma ^{\infty} \D\gamma_1
      (\gamma-\gamma_1)\drond{^2}{{\cal B}^2} \ln S(\gamma_1) 
      \big| _{0}^{{\cal B}}
\label{Mtyp2}                                                 \\
\mean{M _{ee}} 
  &=& \frac{\lambda_0}{\pi}\ \ \ \int_\gamma ^{\infty} \D\gamma_1
      \drond{}{{\cal B}} \ln S(\gamma_1)
\label{Mee}
\:.\eea
When the integrals do not converge at the upper limit, this limit should be 
taken as $1/\tau_e$ where $\tau_e$ is the elastic time.


\section{Definitions and notation\label{DefNot}}

Consider a graph $\cal{G}$ which consists of a set of $V$ vertices linked by 
$B$ bonds. Its 
adjacency matrix $a_{\alpha\beta}$ is a square matrix of size $V$ where 
$a_{\alpha\beta}=1$ if a bond joins the vertices $\alpha$ and 
$\beta$ and $0$ otherwise (in particular $a_{\alpha\alpha}=0$)\footnote{
If a given pair of vertices is linked by $n>1$ bonds, it is always possible to 
introduce additional vertices (see figure \ref{2bonds}), without changing the 
nature of the graph, to go 
back to a situation where there is only one bond between two vertices. 
However, the formalism that is presented in this paper can be easily 
generalized to avoid this procedure and minimize the number of vertices by
suppressing all vertices with coordinence $2$. It is the purpose of appendix
\ref{loops} to discuss this point.}.
The coordinence of vertex $\alpha$ is $m_\alpha=\sum_\beta a_{\alpha\beta}$.
Each bond $(\alpha\beta)$ of length $\lab$ is identified with 
a finite interval $[0,\lab]\in\RR$. The total length of $\cal{G}$ is
$
L=\sum_{{\rm bonds}\ (\alpha\beta)}\lab=
\frac12\sum_{\alpha,\beta}a_{\alpha\beta}\lab
$.
We denote by $x_{\alpha\beta}$ the coordinate on the bond
$(\alpha\beta)$, starting from vertex $\alpha$ (it follows that
$x_{\beta\alpha}=\lab-x_{\alpha\beta}$). A scalar function on $\cal{G}$
is a set of $B$ components functions $\psi_{(\alpha\beta)}(x_{\alpha\beta})$.
When there is no possible ambiguity we will neglect to label components.
The Laplacian on $\cal{G}$ acts as the usual one-dimensional Laplace operator
on each bond:
\be
(\Delta\psi)_{(\alpha\beta)}=
\frac{\D^2}{\D x_{\alpha\beta}^2}\psi_{(\alpha\beta)}(x_{\alpha\beta})
\:.\ee
Its domain is the set of functions that satisfy the following conditions at
the vertices

\noindent ({\it i}) continuity
\be\label{CL1}
\psi_{(\alpha\beta_i)}(x_{\alpha\beta_i}=0)=\psi_\alpha
\ \ \mbox{ for }\ \  i=1,\cdots,m_\alpha
\ee

\noindent ({\it ii}) a second condition sufficient to insure current 
conservation 
\be\label{CL2}
\sum_{i=1}^{m_\alpha}\D_{x_{\alpha\beta_i}}
                     \psi_{(\alpha\beta_i)}(x_{\alpha\beta_i}=0)=0
\ee
where $\D_x\equiv\frac{\D}{\D{x}}$ and the $\beta_i$'s label the $m_\alpha$ 
neighbouring vertices of vertex $\alpha$. The scalar product is defined as
\be
\langle{\psi|\varphi}\rangle=
\sum_{(\alpha\beta)}\int_0^{\lab}\D{x}\,
\psi_{(\alpha\beta)}^*(x)\, \varphi_{(\alpha\beta)}(x)
\:.\ee

In the search of the eigenvalues of the Laplacian, one may construct the 
wave function of energy $E=k^2$ in a way that conditions ({\it i}) is 
satisfied:
$
\psi_{(\alpha\beta)}
=\frac{1}{ \sin kl_{\alpha\beta} }
(\psi_\alpha \sin k(l_{\alpha\beta}-x_{\alpha\beta})
+\psi_\beta\sin k x_{\alpha\beta})
$.
Imposing conditions ({\it ii}) lead to the system of $V$ equations 
$\sum_\beta M_{\alpha\beta}(\gamma=-k^2)\psi_\beta=0$,
where $M$ is given by (\ref{MatrixM1},\ref{MatrixM2}).
The eigenvalues of the Laplacian are solutions of $\det(M(\gamma=-E))=0$
\cite{RudSch53,Ale83,Avr95,KotSmi99}. 

It is useful to introduce some additional notions. 
An {\it arc} $(\alpha\beta)$ is defined as the oriented bond from $\alpha$ 
to $\beta$. Each bond is therefore associated with $2$ arcs. A {\it path} on 
$\cal{G}$ is an ordered sequence of arcs such that the end of each arc 
coincides with the beginning of the one that immediately follows in the 
sequence. Closed paths will play a special role. The set of closed paths
that only differ by a cyclic permutation of the arcs will be called a 
{\it circuit} (or an {\it orbit}). Among all possible circuits we will 
distinguish those that are {\it primitive}. A circuit is said to be primitive 
if it cannot be decomposed as a repetition of any smaller circuit. The 
length of a circuit $C$ will be designated by $l(C)$. 

\begin{figure}[!h]
\begin{center}
\includegraphics{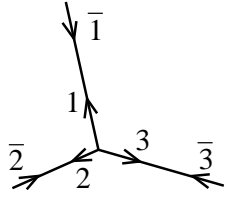}
\end{center}
\caption{A graph with 4 vertices, 3 bonds and 6 arcs.}
\label{exdegraph}
\end{figure}

As an example, let us consider the graph of figure \ref{exdegraph}. 
This graph contains 6 arcs: $1$, $2$, $3$ and the reversed arcs $\bar1$, 
$\bar2$, $\bar3$. 
$(1,\bar1,2,\bar2)$ and $(2,\bar2,1,\bar1)$ are two different paths, two 
possible representations of the same circuit. 
$(1,\bar1,1,\bar1,1,\bar1)$ is not a primitive circuit but 
$(1,\bar1,1,\bar1,1,\bar1,2,\bar2)$ is indeed primitive.


\section{A path integral derivation of the spectral determinant\label{PathInt}}

An expression of the spectral determinant $S(\gamma)=\det(-\Delta+\gamma)$ in 
terms of the determinant of a finite matrix has been derived in 
\cite{PasMon99,Pas98} ({\it cf}. Appendix \ref{DemGilles}).
The purpose of this section is to provide a more direct derivation of this 
result using a path integral formulation.
The spectral determinant may be written as
\be
S(\gamma)=\det(-\Delta+\gamma)=\left(
  \int_{\phi\rm\:on\:Graph}\dphi
  \EXP{-\frac12\int_{\rm Graph}\bar\phi(-\Delta+\gamma)\phi}
\right)^{-1}
\ee
where $\phi$ is a complex field defined on the graph ${\cal G}$. The path 
integral is performed over all fields satisfying the conditions
(\ref{CL1},\ref{CL2}) and the integral in the exponential along all branches 
of ${\cal G}$. 
Since the field is continuous at each vertex the
path integral may be decomposed on each bond $(\alpha\beta)$ as an
integration over fields that take fixed values $\phi_\alpha$ and
$\phi_\beta$ on the two sides of the bond
\be\label{pi1}
S(\gamma)^{-1}=\int\prod_{{\rm vertices}\:\alpha}
\D\phi_\alpha\D\bar\phi_\alpha
\prod_{\stackrel{\rm bonds}{(\alpha\beta)}}
\int_{\phi(0)=\phi_\alpha}^{\phi(\lab)=\phi_\beta}
\dphi\EXP{-\frac12\int_0^{\lab}\D{x}\,\bar\phi(x)(-\D_x^2+\gamma)\phi(x)}
\ee
where $\D\phi\D\bar\phi=\D\re\phi\,\D\im\phi$.
This involves, after an integration by parts, the following quantity\footnote{
The notation $\sum_{(\alpha\beta)}\cdots$ and $\prod_{(\alpha\beta)}\cdots$
means, throughout this paper, a summation or a product over bonds 
(and not arcs).}
\bea
&&\prod_{(\alpha\beta)}
\int_{\phi(0)=\phi_\alpha}^{\phi(\lab)=\phi_\beta}
\dphi\EXP{\frac12\bar\phi\D_{x}\phi\big|_0^{\lab}}
\EXP{-\frac12\int_0^{\lab}\D{x}\,(|\D_x\phi|^2+\gamma|\phi|^2)}
\nonumber\\ \label{pi2}
&&=\exp\Big({-\frac12\sum_{\alpha=1}^{V}\bar\phi_\alpha
         \sum_{i=1}^{m_\alpha}\D_{x_{\alpha\beta_i}}\phi
         ({x_{\alpha\beta_i}=0})}
       \Big)
\prod_{(\alpha\beta)}
\int_{\phi(0)=\phi_\alpha}^{\phi(\lab)=\phi_\beta}
\dphi
\EXP{-\frac12\int_0^{\lab}\D{x}\,(|\D_x\phi|^2+\gamma|\phi|^2)}
\:.\eea
For the sum of boundary terms in the exponential we have replaced the sum 
over the bonds by a sum over the vertices, a trick that will be frequently
used throughout the rest of this paper.
Equation (\ref{CL2}) implies that the boundary terms vanish.
The above path integral, involving a zero-dimensional Gaussian field theory,
can easily be integrated out using standard quantum mechanical
techniques. On each bond the path integral can be expressed in terms of
the propagator of a two-dimensional harmonic oscillator of frequency
$\omega=\sqrt{\gamma}$
\be
{\cal N}_r^{-1}
\int_{\phi(0)=\phi_\alpha}^{\phi(\lab)=\phi_\beta}
\dphi
\EXP{-\frac12\int_0^{\lab}\D{x}\,(|\D_x\phi|^2+\gamma|\phi|^2)}
=G_{\lab}^{\omega=\sqrt{\gamma}}(\vec\phi_\beta,\vec\phi_\alpha)=
\bra{\vec\phi_\beta}\EXP{-\frac{\lab}{2}
\left(-\vec\nabla_\phi^2+\gamma\vec\phi^2\right)}
\ket{\vec\phi_\alpha}
\:,\ee
where $\vec\phi=(\re\phi,\im\phi)$ and ${\cal N}_r$ is a constant
independent of $\gamma$ depending on the precise normalization chosen in the
definition of the path integral, or in other terms on the choice of
regularization made to define the determinant. In the following we will drop
this inessential normalization constant.
Using the expression of the propagator \cite{FeyHib65}
\be
G^{\omega}_{x}(\vec\phi,\vec\phi\,')=
\frac{\omega}{2\pi\sinh(\omega x)}
\EXP{-\frac{\omega}{2\sinh(\omega x)}(
     \cosh(\omega x)(\vec\phi^2+\vec\phi\,'^2)-2\vec\phi\cdot\vec\phi\,')}
\ee
one may re-scale the fields $\phi_\alpha$ to extract a $\gamma$-dependent
factor:
\be\label{sp}
S(\gamma)^{-1}=\gamma^{\frac{B-V}{2}}
\int \prod_{\alpha=1}^V\D\phi_\alpha\D\bar\phi_\alpha
\prod_{(\alpha\beta)}
G_{\sqrt{\gamma}\lab}(\vec\phi_\beta,\vec\phi_\alpha)
\:,\ee
where 
$G_{x}(\vec\phi,\vec\phi\,')\equiv G^{\omega=1}_{x}(\vec\phi,\vec\phi\,')$.
This expression will be used as a starting point for the derivation of the
following section.
Using the $V\times V$ matrix $M$ introduced in section \ref{Intro} 
\bea\label{M1}
M_{\alpha\alpha}&=&\sum_{i=1}^{m_\alpha}\coth(\sqrt{\gamma}l_{\alpha\beta_i}) 
\\
\label{M2}
M_{\alpha\beta} &=&-\frac{1}{\sinh(\sqrt{\gamma}\lab)}
\ \ \mbox{ if $(\alpha\beta)$ is a bond} \\
                &=& 0 \ \ \mbox{ otherwise}\nonumber
\:,\eea
we may express the determinant as
\be
S(\gamma)^{-1}=\gamma^{\frac{B-V}{2}}
\prod_{(\alpha\beta)}\frac{1}{2\pi\sinh(\sqrt{\gamma}\lab)}
\int \Big(\prod_{\alpha=1}^V\D\phi_\alpha\D\bar\phi_\alpha\Big)
\EXP{-\frac{1}{2}\sum_{\alpha,\beta}\bar\phi_\alpha M_{\alpha\beta}\phi_\beta}
\:.\ee
After performing the integration over $\phi_\alpha$ one finds
\be\label{MonPas}
S(\gamma)=\left(\frac{\sqrt{\gamma}}{2\pi}\right)^{V-B}
\prod_{(\alpha\beta)}\sinh(\sqrt{\gamma}\lab)
\ \det(M)
\ee
which is the expression given in \cite{PasMon99} (see appendix 
\ref{DemGilles}) up to an inessential numerical factor $(2\pi)^{B-V}$ that 
will be dropped in the following.


\section{The spectral determinant in terms of the link matrix\label{Jean}}

Equation (\ref{MonPas}) expresses the spectral determinant as the
determinant of a finite $V\times V$ matrix. The matrix $M$ describes the
topology of the graph, telling us which vertices are coupled, and also
contains the metric information. At this stage, the metric and topological
information are inextricably mixed. 
The purpose of this section is to derive another
expression of the spectral determinant in terms of the determinant of a  
$2B\times2B$ matrix whose natural basis is the set of arcs (oriented bonds).
An advantage of this formulation is that it leads to an expansion of the 
spectral determinant as a sum of terms that will be interpreted in the next 
section as the contribution of periodic orbits.

Our starting point is expression (\ref{sp}). Since the $\gamma$
dependence of the determinant is simple, one may set $\gamma=1$ and recover 
the $\gamma$ dependence at the end. 
The first step is to find a more convenient expression of the propagator in
(\ref{sp}).
To begin let us consider the one-dimensional harmonic oscillator propagator
$g_t(x,x')=\bra{x}\EXP{-\frac{t}{2}(-\D_x^2+x^2)}\ket{x'}$. This
propagator may be expanded over the eigenstates
$\varphi_n(x)=\frac{1}{\pi^{1/4}\sqrt{2^n n!}}H_n(x)\EXP{-\frac12x^2}$ of
energies $E_n=n+\frac12$, where $H_n(x)$ are Hermite polynomials.
Using the generating function of Hermite polynomials
$\sum_{n=0}^{\infty}H_n(x)\frac{\lambda^n}{n!}=\EXP{2\lambda x-\lambda^2}$,
the propagator may be rewritten as
\be
g_{\lab}(x_\alpha,x_\beta)=\frac{\EXP{-\frac{\lab}{2}}}{\sqrt\pi}
\EXP{-\frac12(x_\alpha^2+x_\beta^2)}
\sum_{n=0}^\infty\frac{\EXP{-n\,\lab}}{2^n\,n!}
  \partial_{\lamab}^n\EXP{2x_\alpha\lamab-\lamab^2}
  \partial_{\lamab}^n\EXP{2x_\beta\lamba-\lamba^2}
  \big|_{\lamab,\lamba=0}
\:.\ee
A re-summation of the series gives
\be
g_{\lab}(x_\alpha,x_\beta)=\frac{\EXP{-\frac{\lab}{2}}}{\sqrt\pi}
\EXP{-\frac12(x_\alpha^2+x_\beta^2)}
\EXP{\frac12\EXP{-\lab}\partial_{\lamab}\partial_{\lamba}}
\EXP{2x_\alpha\lamab+2x_\beta\lamba-\lamab^2-\lamba^2}\big|_{\lamab,\lamba=0}
\:.\ee
Writing the two-dimensional propagator as a product of one-dimensional
propagators
$
G_{\lab}(\vec r_\alpha,\vec r_\beta)=
g_{\lab}(x_\alpha,x_\beta)g_{\lab}(y_\alpha,y_\beta)$
one is led to an analogous expression for
$G_{\lab}(\vec r_\alpha,\vec r_\beta)$. For each bond
one has to introduce two couples of variables $(\lamab,\lamba)$ and
$(\lamab',\lamba')$ associated to $x_\alpha,x_\beta$ and
$y_\alpha,y_\beta$ respectively. Using the complex notations
$z_\alpha=x_\alpha+\I y_\alpha$ and $\Lamab=\lamab+\I\lamab'$ (one recalls
that $\dz=\frac12(\partial_x-\I\partial_y)$) one eventually finds
\be\label{propa}
G_{\lab}(z_\alpha,z_\beta)=
\frac{\EXP{-\lab}}{\pi}
\EXP{-\frac12(|z_\alpha|^2+|z_\beta|^2)} 
\Oab\EXP{ z_\alpha\bar\Lamab+\bar z_\alpha\Lamab
         +z_\beta\bar\Lamba+\bar z_\beta\Lamba-|\Lamab|^2-|\Lamba|^2}
\:,\ee
where $\Oab$ is the following operator:
\be
\Oab=
\exp\Big(
      {\EXP{-\lab}
      ( \partial_{\Lamab}\partial_{\bar\Lamba}
       +\partial_{\bar\Lamab}\partial_{\Lamba})}
    \Big)\Big|_{\Lamab,\Lamba=0}
\:.\ee
Equation (\ref{propa}) shows that one has to introduce two variables $\Lamab$ 
and $\Lamba$ per propagator in (\ref{sp}), {\it i.e.} each arc $(\alpha\beta)$
is associated with a $\Lamab$.

From (\ref{sp},\ref{propa}) it follows that
\bea
S(\gamma=1)^{-1}&=&2^{B-V}\frac{\EXP{-L}}{\pi^V}\ 
{\cal O}\bigg[\prod_{(\alpha\beta)}\left(\EXP{-|\Lamab|^2-|\Lamba|^2}\right)
\nonumber\\
&&
\int\prod_{\alpha}\left(\D{\phi_\alpha}\D{\bar\phi_\alpha}\,
\EXP{-\frac12m_\alpha|\phi_\alpha|^2
    +\sum_{i=1}^{m_\alpha}(\phi_\alpha\bar\Lambda_{\alpha\beta_i}+
                           \bar\phi_\alpha\Lambda_{\alpha\beta_i})}
\right)\bigg]
\label{SJ1}
\eea
where the operator ${\cal O}=\prod_{(\alpha\beta)}\Oab$ is understood as
acting on both terms that follow it (one has multiplied (\ref{sp}) by the
numerical factor $(2\pi)^{V-B}$ to make it disappear in the final result). 
Again, one has replaced in (\ref{sp})
the product over bonds by a product over vertices.
Integration over $\phi_\alpha$ is then straightforward and leads to 
\be
S(\gamma=1)^{-1}=\frac{2^B\EXP{-L}}{\prod_{\alpha}m_\alpha}
\ {\cal O}\EXP{\Lambda^\dagger{Q}\Lambda}
\ee
where $\Lambda$ is the $2B$-component complex vector:
\be
\Lambda=
\left(\begin{array}{c}
\Lambda_{1\alpha_1}\\
\vdots\\
\Lambda_{1\alpha_{m_1}}\\
\hline
\Lambda_{2\beta_1}\\
\vdots\\
\Lambda_{2\beta_{m_2}}\\
\hline
\vdots
\end{array}\right)
\ee
and
\be
\Lambda^\dagger=
\left(\begin{array}{ccc|ccc|c}
\bar\Lambda_{1\alpha_1} & \cdots & \bar\Lambda_{1\alpha_{m_1}} & 
\bar\Lambda_{2\beta_1}  & \cdots & \bar\Lambda_{2\beta_{m_2}}  & \cdots
\end{array}\right)
\:,\ee
in which the $\Lamab$ are grouped by vertices. The first group
$(\Lambda_{1\alpha_1},\cdots,\Lambda_{1\alpha_{m_1}})$ is related to the 
$m_1$ arcs $\alpha_1,\alpha_2,\cdots,\alpha_{m_1}$
issuing from vertex $\alpha=1$. In this basis, $Q$ is a $2B\times2B$ block 
diagonal matrix (with $V$ blocks); it only couples arcs that start from the 
same vertex
\be\label{Q}
Q=
\left(\begin{array}{c|c|c|c|c}
Q_1    & 0      & 0      & \cdots & 0      \\
\hline
0      & Q_2    & 0      & \cdots & 0      \\
\hline
0      & 0      & Q_3    & \cdots & 0      \\
\hline
\vdots & \vdots & \vdots & \ddots & \vdots \\
\hline
0      & 0      & 0      & \cdots & Q_V
\end{array}\right)
\ee
where each $m_\alpha\times m_\alpha$ sub-matrix $Q_\alpha$ attached
to a given vertex $\alpha$ has $\frac{2}{m_\alpha}-1$ on its diagonal and
$\frac{2}{m_\alpha}$ everywhere else. Using a tensorial notation, one may
write the matrix element of $Q$ between the two arcs $(\alpha\beta)$ and
$(\mu\nu)$ as
\be
Q_{(\alpha\beta)(\mu\nu)}=a_{\alpha\beta}a_{\mu\nu}
\delta_{\alpha\mu}\left(\frac{2}{m_\alpha}-\delta_{\beta\nu}\right)
\:,\ee
where $\delta_{\alpha\beta}$ is the Kronecker symbol; the connectivity
matrices $a_{\alpha\beta}$ and $a_{\mu\nu}$ are here to recall that
$(\alpha\beta)$ and $(\mu\nu)$ are arcs.

The action of ${\cal O}$ on $\EXP{\Lambda^\dagger{Q}\Lambda}$ is
complicated because the argument of the exponential is quadratic in the
$\Lamab$'s. This action would become much more simple on the exponential of a
linear form; this can be achieved by introducing an additional integration
over a $2B$-component complex vector $W$
\be
S(\gamma=1)^{-1}=
\frac{2^B\EXP{-L}}
     {\left(\prod_{\alpha}m_\alpha\right)\pi^{2B}}\det(Q^{-1})
\int\D{W^\dagger}\D{W}\,\EXP{-W^\dagger{Q}^{-1}W}
{\cal O}\EXP{\Lambda^\dagger W+W^\dagger\Lambda}
\:.\ee
The action of ${\cal O}$ is now simple by noticing that 
$
\partial_{\Lamab}\partial_{\bar\Lamba}
\EXP{\Lambda^\dagger W+W^\dagger\Lambda}
=\bar w_{\alpha\beta}w_{\beta\alpha}\EXP{\Lambda^\dagger
W+W^\dagger\Lambda}$.
To write the action of ${\cal O}$ on
$\EXP{\Lambda^\dagger W+W^\dagger\Lambda}$, one has to replace 
$\partial_{\Lamab}$ by $\bar w_{\alpha\beta}$ and 
$\partial_{\bar\Lamab}$ by $w_{\alpha\beta}$. One finally ends up with
\be
S(\gamma=1)^{-1}=
\frac{2^B\EXP{-L}}
     {\left(\prod_{\alpha}m_\alpha\right)\pi^{2B}}
\det(Q^{-1}) \int\D{W^\dagger}\D{W}\,\EXP{-W^\dagger{Q}^{-1}W+W^\dagger RW}
\:,\ee
where it is clear from the expression of ${\cal O}$ that the matrix $R$ only
couples the arc $(\alpha\beta)$ to the time reversed arc $(\beta\alpha)$.
Thus, $R$ may be written
\be
R_{(\alpha\beta)(\mu\nu)}=a_{\alpha\beta}a_{\mu\nu}
\delta_{\alpha\nu}\delta_{\beta\mu}\EXP{-\lab}
\:.\ee
It is then straightforward to perform the Gaussian integration and get
the final expression for the spectral determinant 
\be\label{resJean}
S(\gamma)=\gamma^{\frac{V-B}{2}}\EXP{\sqrt{\gamma}\,L}
\Big(\prod_{\alpha}m_\alpha\Big)2^{-B}\ \det(1-E)
\:.\ee
where we have introduced $E=(QR)^{\rm T}$ 
(the transposition is defined as 
$(E^{\rm T})_{(\alpha\beta)(\mu\nu)}=E_{(\mu\nu)(\alpha\beta)}$).

Using this tensorial notation  the product $QR$ is particularly simple to 
perform
\be
E_{(\alpha\beta)(\mu\nu)}=a_{\alpha\beta}a_{\mu\nu}
\delta_{\beta\mu}\left(\frac{2}{m_\beta}-\delta_{\alpha\nu}\right)
\EXP{-\sqrt{\gamma}\,l_{\alpha\beta}}
\:.\ee
The spectral determinant is now given by the determinant of a $2B\times2B$
matrix $E$ whose basis consists of the set of arcs. Rewriting the matrix
$E=D\epsilon$ one may clearly separate a part $\epsilon$ that only depends on
the topology of the graph 
\be\label{mateps}
\epsilon_{(\alpha\beta)(\mu\nu)}=a_{\alpha\beta}a_{\mu\nu}
\delta_{\beta\mu}\left(\frac{2}{m_\beta}-\delta_{\alpha\nu}\right)
\ee
and a diagonal part $D$ depending on the metric properties
\be
D_{(\alpha\beta)(\mu\nu)}=a_{\alpha\beta}a_{\mu\nu}
\delta_{\alpha\mu}\delta_{\beta\nu}\EXP{-\sqrt{\gamma}\,l_{\mu\nu}}
\:.\ee

\bigskip

We end this section by mentioning the physical meaning of the
different matrices that appear in the derivation given above. 

In order to
simplify slightly our notations arcs will now be labeled by Roman letters
$i,j,\ldots$ ($i\equiv(\alpha\beta)$). The matrix $E$ is then expressed as
$E_{ij}=\epsilon_{ij}\EXP{-\sqrt\gamma l_i}$. 
Our $\epsilon_{ij}$ coincides with
the one introduced in \cite{Rot83}: $\epsilon_{ij}=\frac{2}{m_\alpha}$ if the 
end of arc $i$ and the beginning of arc $j$ are the vertex $\alpha$ of 
coordinence $m_\alpha$, $\epsilon_{ij}=\frac{2}{m_\alpha}-1$ if moreover the 
two arcs are time-reversed, and $\epsilon_{ij}=0$ otherwise. 

\begin{figure}[!h]
\begin{center}
\includegraphics{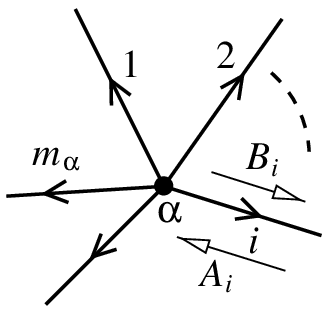}
\end{center}
\caption{}
\label{vertexscatt}
\end{figure}

Following Kottos and Smilansky \cite{KotSmi99},
one may take a scattering point of view to construct the wave function on 
the graph. Let us consider first the set of arcs $i\in\{1,...,m_\alpha\}$ 
starting at vertex $\alpha$. On each arc $i$ one writes the wave function as 
the superposition of an incoming and an outcoming plane wave (see figure 
\ref{vertexscatt}):
\be\label{scatstate}
\psi_i(x)=A_i\EXP{-\I kx}+B_i\EXP{\I kx}
\:.\ee 
The scattering matrix
$s_\alpha$ at the vertex is defined as the matrix relating incoming to
outcoming amplitudes: $B_i=\sum_{j}(s_\alpha)_{ij}A_j$ for 
$i,j\in\{1,...,m_\alpha\}$. Conditions (\ref{CL1},\ref{CL2}) lead to the 
following result: $s_\alpha=Q_\alpha$. On the graph, the $2B$ outcoming
amplitudes are then coupled to the $2B$ incoming amplitudes by \cite{KotSmi99}
\be
B_i=\sum_{j}Q_{ij}A_j
\ee 
(matrix $Q$ indeed couples arcs beginning at same vertex). The physical 
meaning of the matrix $Q$ is now clear: it describes the scattering by the 
$V$ vertices. As in section \ref{DefNot}
we use the notation $\bar i$ for the time-reversed arc of $i$. Amplitudes
of two time-reversed arcs are related by the two obvious relations:
$B_i=A_{\bar i}\EXP{-\I kl_i}$ and $A_i=B_{\bar i}\EXP{\I kl_i}$
(due to the fact that $\psi_i(x)=\psi_{\bar i}(l_i-x)$).
Then one has $B_i=\sum_{j}Q_{ij}\EXP{\I kl_j}B_{\bar j}$. Using the fact that
$Q_{\bar ij}=\epsilon_{ij}$ one finds that 
\be
B_{\bar i}=\sum_{j}E_{ij}(\gamma=-k^2)B_{\bar j}
\:.\ee
Then the state (\ref{scatstate}) of energy $k^2$ belongs to the spectrum if 
$\det(1-E(\gamma=-k^2))=0$ \cite{Car99}. As equation 
$\det(M(\gamma=-k^2))=0$, this equation gives the spectrum of the Laplacian 
but not the $\gamma$-dependent pre-factor of $S(\gamma)$.


\section{A trace formula \label{TraFor}}

The result (\ref{resJean}) is particularly suitable to derive a trace
formula that expresses the determinant in terms of
contributions of closed trajectories (orbits) on the graph. 
Again we may set 
$\gamma=1$ for the sake of simplicity. We may expand the determinant 
\be\label{ldet}
\ln\det(1-E)=-\sum_{n=1}^\infty\frac{1}{n}\tr{E^n}
\ee
and write the trace as a sum of terms
\be\label{tracebrute}
\tr{E^n}=\sum_{i_1,\cdots,i_n}
\epsilon_{i_1i_2}\epsilon_{i_2i_3}\cdots\epsilon_{i_ni_1}
\EXP{-(l_{i_1}+\cdots+l_{i_n})}
\:,\ee
each of which is associated with a sequence of arcs that forms a closed path
on the graph. Following the notations of \cite{Rot83} one may attach the
quantity 
\be
\alpha(C_n)=\epsilon_{i_1i_2}\epsilon_{i_2i_3}\cdots\epsilon_{i_ni_1}
\ee
to the orbit (or circuit) $C_n=(i_1,i_2,\cdots,i_n)$ composed of $n$ arcs,
and denote its length by $l(C_n)=l_{i_1}+\cdots+l_{i_n}$.
Each term of $\tr{E^n}$ corresponds to a path associated with either a
primitive orbit of $n$ steps or $k$ repetitions of a primitive orbit of
$p$ steps such as $n=k\,p$.
Primitive orbits will be labeled as $\tilde C$.
A given primitive orbit of $n$ steps appears $n$ times in $\tr{E^n}$,
corresponding to the $n$ cyclic permutations of the indices that give the
same orbit on the graph.
On the other hand the term associated to $k$ repetitions of a
primitive orbit $\tilde C_p$ appears $p$ times corresponding to the $p$
different permutations of the indices that appear in $\tr{E^n}$.
Using these remarks, instead of summing over all paths of length $n$ as in
(\ref{tracebrute}), one may sum over the $k$-repetition of primitive
orbits of length $p$ provided that $n=k\,p$:
\be
\tr{E^n}=\sum_{k, \tilde C_p\:{\rm with}\:n=kp} 
p\ \alpha(\tilde C_p)^k\EXP{-k\,l(\tilde C_p)}
\:,\ee
we have used the fact that if $C_n$ is a $k$-repetition of $\tilde C_p$
then $\alpha(C_n)=\alpha(\tilde C_p)^k$ and $l(C_n)=k\,l(\tilde C_p)$.
Introducing this expression in (\ref{ldet}) gives
\be
\ln\det(1-E)=-\sum_{\tilde C_p}\sum_{k=1}^\infty\frac{1}{kp}
p\ \alpha(\tilde C_p)^k\EXP{-k\,l(\tilde C_p)}
\:,\ee
Instead of summing over $n$, $k$ and $\tilde C_p$ with the constraint
$kp=n$, one may sum over all primitive orbits $\tilde C_p$ and their
$k$-repetition independently. This immediately leads \cite{Eck93} to 
\be
\ln\det(1-E)=\sum_{\tilde C_p}
\ln\left(1-\alpha(\tilde C_p)\EXP{-l(\tilde C_p)}\right)
\:.\ee
One thus gets an expression of the spectral determinant in terms of an
infinite product over all the primitive orbits $\tilde C$, in infinite
number\footnote{Note however there are two trivial graphs for which one can
construct only one primitive orbit: the circle and the line.}, that can be 
constructed on the graph: 
\be\label{Sip}
S(\gamma)=\gamma^{\frac{V-B}{2}}\EXP{\sqrt{\gamma}\,L}
\Big(\prod_{\alpha}m_\alpha\Big)2^{-B}
\prod_{\tilde C}\left(
  1-\alpha(\tilde C)\EXP{-\sqrt{\gamma}\,l(\tilde C)}\right)
\:.\ee

A similar trace formula for the partition function was first derived by Roth
\cite{Rot83}. It is indeed possible to recover it by starting from
(\ref{Sip}). The log-derivative of the spectral determinant is equal to 
the Laplace transform of the partition function (\ref{relZS}). 
From (\ref{Sip}) one gets
\be
\drond{}{\gamma}\ln S(\gamma)=\frac{L}{2\sqrt\gamma}+\frac{V-B}{2\gamma}
+\frac{1}{2\sqrt\gamma}\sum_{\tilde C}l(\tilde C)\sum_{k=1}^\infty
\alpha(\tilde C)^k\EXP{-k\sqrt{\gamma}\,l(\tilde C)}
\:,\ee
where one has expanded the denominator coming from the derivative of the
logarithm that appears writing $\ln S(\gamma)$. It is then possible to
group the two sums in a unique sum over all orbits $C$, including the
repetitions of the primitive orbits
\be
\drond{}{\gamma}\ln S(\gamma)=\frac{L}{2\sqrt\gamma}+\frac{V-B}{2\gamma}
+\frac{1}{2\sqrt\gamma}\sum_{C} l(\tilde C)
\alpha(C)\EXP{-\sqrt{\gamma}\,l(C)}
\:.\ee
In this summation, $\tilde{C}$ designates the primitive orbit associated 
with a given orbit $C$.

Using the following identity 
$
\frac{1}{2\sqrt\gamma}\EXP{-\sqrt{\gamma}\,l}=\frac{1}{2\sqrt\pi}
\int_0^\infty\D{t}\,\frac{1}{\sqrt{t}}\EXP{-\gamma t-\frac{l^2}{4t}}
$ 
it is then easy to extract the inverse Laplace transform of the previous
expression and recover the trace formula first obtained by Roth
\cite{Rot83,Rot83a}
\be\label{Roth}
Z(t)=\frac{L}{2\sqrt{\pi t}}+\frac{V-B}{2}+\frac{1}{2\sqrt{\pi t}}
\sum_{C} l(\tilde C)\alpha(C)\EXP{-\frac{l(C)^2}{4t}}
\:.\ee

Let us notice that if one considers a vertex $\alpha$ of coordinence 
$m_\alpha=2$, the circuits that contain a reflection at this vertex have
weights $\alpha(C)=0$, therefore such circuits should not be considered in 
the expansion. Since one can always introduce an arbitrary number of vertices
on any bond without changing the properties of the graph, this remark 
insures that the number of orbits will not vary doing so. Moreover
this shows that a graph can always be simplified to minimize $B$ and $V$ by 
suppressing all vertices of coordinence $2$.


\section{Simplification of the trace formula - Diagrammatic Expansion of the 
         spectral determinant \label{DiagExp}}

In the previous section one has expressed the determinant $\det(1-E)$, which
can be expanded to give a finite number of terms, as an infinite product 
(\ref{Sip}). The purpose of this section is to show how the infinite product 
eventually simplifies to give a finite number of terms. 
One interest of this discussion is to provide a diagrammatic method
to construct systematically the different terms of $S(\gamma)$.

We will consider the quantity that appears in (\ref{Sip}):
\be
\hat{S}=
\prod_{\tilde C}\left(
  1-\alpha(\tilde C)\EXP{-l(\tilde C)}\right)
\:.\ee
The generalization of the following discussion to the case with a magnetic 
field is straightforward (see (\ref{Sipwf})).

We will call $-\alpha(C)\EXP{-l(C)}$ the weight of the orbit. We associate
to each term in the expansion of $\hat{S}$ a diagram, that represents
either the contribution of an orbit or the product of such contributions; in
the last case the diagram represents the superposition of the different
orbits.

\begin{figure}[!h]
\begin{center}
\includegraphics{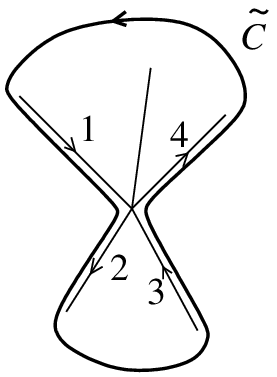}
\hspace{2cm}
\includegraphics{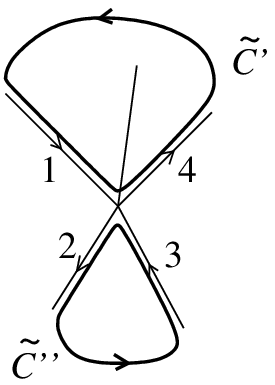}
\end{center}
\caption{}
\label{2diag}
\end{figure}

To understand how the simplifications occur let us consider a primitive
orbit $\tilde{C}$ that passes twice through the same vertex without using
the same arcs as on figure \ref{2diag}.
Using the fact that
$\epsilon_{12}=\epsilon_{34}=\epsilon_{14}=\epsilon_{32}$ it is easy to see
that
\be
-\alpha(\tilde{C})\EXP{-l(\tilde{C})}=
- \left(-\alpha(\tilde{C}')\EXP{-l(\tilde{C}')}\right)
\:\left(-\alpha(\tilde{C}'')\EXP{-l(\tilde{C}'')}\right)
\ee
where $\tilde{C}'$ and $\tilde{C}''$ are obtained by crossing the paths 
at the vertex as shown in figure \ref{2diag}. The consequence is
that in the expansion of $\hat{S}$, the product of the weights of orbits
$\tilde{C}'$ and $\tilde{C}''$ cancels with the contribution of $\tilde{C}$. 

\begin{figure}[!h]
\begin{center}
\includegraphics[scale=0.75]{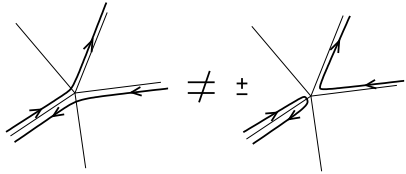}
\end{center}
\caption{Take care not to introduce some reflection at a vertex using the
rules (\ref{rules}).}
\label{Verbotten}
\end{figure}

These kind of relations may be represented diagrammatically as 
\be\label{rules}
\diagram{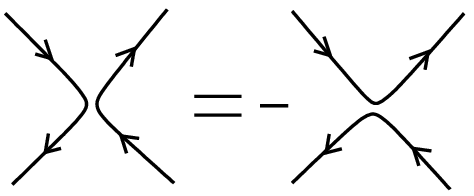}{0.75}{-0.5cm}
\hspace{1cm}{\rm and}\hspace{1cm}
\diagram{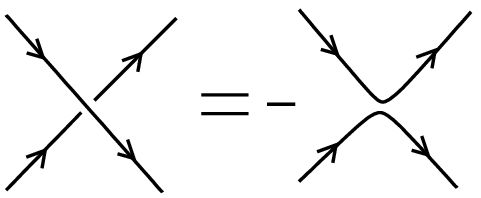}{0.75}{-0.5cm}
\ee
Using these relations one must take care of not introducing some reflection at
a vertex like in figure \ref{Verbotten}. Indeed the diagram on the left of
figure \ref{Verbotten} has a weight proportional to 
$(\frac{2}{m_\alpha})^2$ whereas the diagram on the right has a
weight proportional to 
$-(\frac{2}{m_\alpha}-1)\frac{2}{m_\alpha}$.

Note that the case of a vertex with coordinence $m_\alpha=4$ is special and
may bring some additional rules which simplify the expansion of $\hat{S}$.
Those two additional rules are:
\be
\diagram{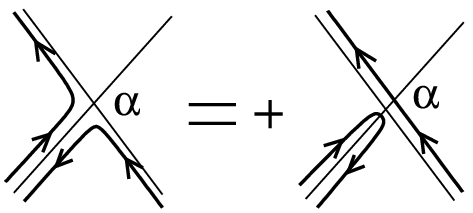}{0.75}{-0.5cm}
\ \ \mbox{ for }m_\alpha=4
\ee
and
\be
\diagram{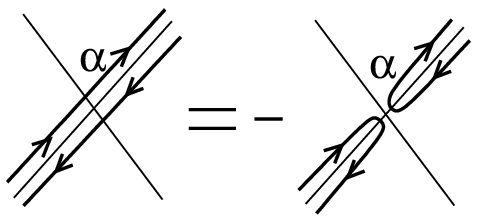}{0.75}{-0.5cm}
\ \ \mbox{ for }m_\alpha=4
\ee

What kind of consequences can be deduced from these rules (\ref{rules})~? 
Let us consider a primitive orbit that contains twice the same arc. This 
orbit may be factorised using the de-crossing rule (\ref{rules}):
\be
\diagram{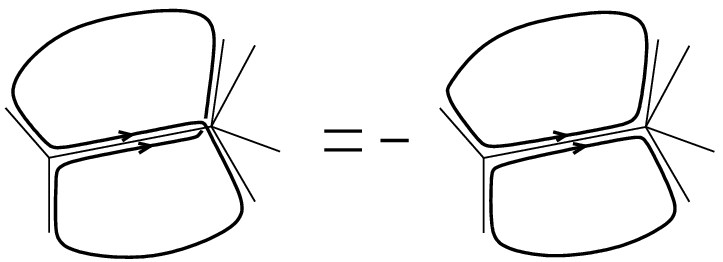}{1}{-1cm}
\:.\ee
This equation implies that one should not include in the expansion of
$\hat{S}$ the diagrams that contain more than once a given arc. This remark
implies that the number of diagrams to be considered is finite.
In the diagrams that remain in the expansion, a bond appears in the orbits
at most twice, corresponding to the two reversed arcs; a by-product of this
remark is that the longest orbits that can be constructed 
satisfying this rule have lengths $2L$.

\subsection*{Example of a diagrammatic expansion}

\begin{figure}[!h]
\begin{center}
\includegraphics{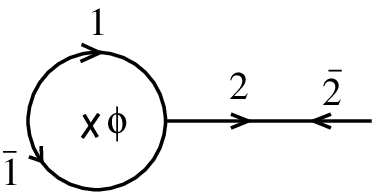}
\end{center}
\caption{}
\label{ouvrebouteille}
\end{figure}

As an example, consider the diagram of figure \ref{ouvrebouteille}, which
consists of an arm connected to a ring pierced by a flux $\phi$; 
this geometry was considered for the study of persistent currents in the
case of a one-channel clean ring \cite{But85} and in the case of a metallic
diffusive ring \cite{PasMon99}.  
The first step is to construct all the periodic orbits that will be involved
in the expansion (see the table \ref{primorb}). One considers the situation 
where the ring is pierced by a
flux $\phi$ to distinguish between time-reversed orbits;
$\theta=2\pi\phi/\phi_0$ where $\phi_0=h/e$ is the flux quantum. We call $l$
the perimeter of the ring and $b$ the length of the arm. As an example let us
compute the $\alpha(C)$ coefficient for the last orbit of the table 
\ref{primorb}
$\tilde{C}=(1,2,\bar2,\bar1)$: 
$\alpha(\tilde{C})=
\epsilon_{12}\epsilon_{2\bar2}\epsilon_{\bar2\bar1}\epsilon_{\bar11}$
with $\epsilon_{\bar11}=-\frac13$,
$\epsilon_{12}=\epsilon_{\bar2\bar1}=\frac23$ and $\epsilon_{2\bar2}=1$.

\begin{table}[!h]
\begin{center}
\begin{tabular}{|c|c|}
\hline
primitive orbit $\tilde{C}$ 
        & weight $-\alpha(\tilde{C})\EXP{-l(\tilde{C})+\I\theta(\tilde{C})}$\\
\hline\hline
\diagram{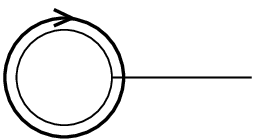}{1}{-0.5cm} & $-\frac23\EXP{-l-\I\theta}$    \\
\diagram{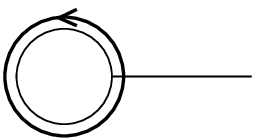}{1}{-0.5cm} & $-\frac23\EXP{-l+\I\theta}$    \\
\diagram{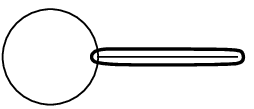}{1}{-0.5cm} & $\frac13\EXP{-2b}$             \\
\diagram{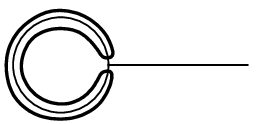}{1}{-0.5cm} & $-\frac19\EXP{-2l}$            \\
\diagram{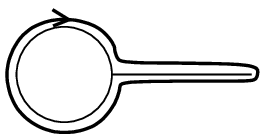}{1}{-0.5cm} & $-\frac49\EXP{-l-2b-\I\theta}$ \\
\diagram{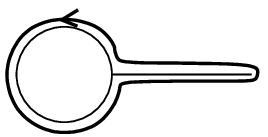}{1}{-0.5cm} & $-\frac49\EXP{-l-2b+\I\theta}$ \\
\diagram{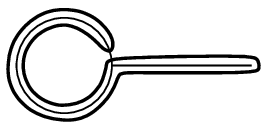}{1}{-0.5cm} & $\frac{4}{27}\EXP{-2l-2b}$     \\
\diagram{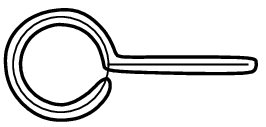}{1}{-0.5cm} & $\frac{4}{27}\EXP{-2l-2b}$     \\
\hline
\end{tabular}
\end{center}
\caption{\label{primorb}}
\end{table}


The construction of all diagrams of the expansion requires to combine all
these orbits provided that the resulting diagrams do not contain twice the
same arc. This leads to
\bea\label{expansion}
\hat{S}=1 
+ \bigg( \diagram{po1.eps}{0.75}{-0.3cm}
        +\diagram{po2.eps}{0.75}{-0.3cm} \bigg)
+ \bigg( \diagram{po3.eps}{0.75}{-0.3cm} \bigg) 
+ \bigg( \diagram{po4.eps}{0.75}{-0.3cm}
        +\diagram{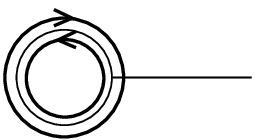}{0.75}{-0.3cm} \bigg) \nonumber\\
+ \bigg( \diagram{po5.eps}{0.75}{-0.3cm}
        +\diagram{po6.eps}{0.75}{-0.3cm}
        +\diagram{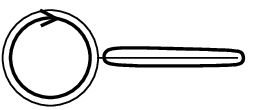}{0.75}{-0.3cm}
        +\diagram{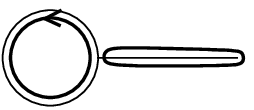}{0.75}{-0.3cm} \bigg)
+ \bigg( \diagram{po7.eps}{0.75}{-0.3cm}
        +\diagram{po8.eps}{0.75}{-0.3cm}  \nonumber\\
        +\diagram{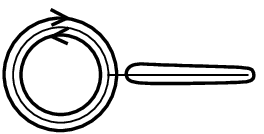}{0.75}{-0.3cm} 
        +\diagram{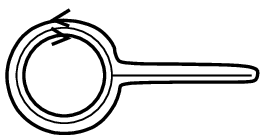}{0.75}{-0.3cm}
        +\diagram{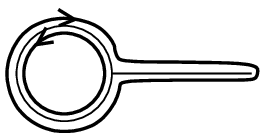}{0.75}{-0.3cm}
        +\diagram{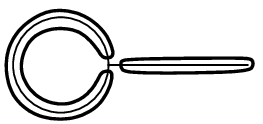}{0.75}{-0.3cm} \bigg)
\eea
where it is understood that a diagram with two primitive orbits is equivalent
to the product of diagrams for each orbit, as said above. For example
\be
\diagram{po16.eps}{0.75}{-0.3cm}\equiv
\diagram{po1.eps}{0.75}{-0.3cm}\times
\diagram{po6.eps}{0.75}{-0.3cm}
\:.\ee
Using the weights of the orbits given in the table one eventually finds
\bea
\hat{S}&=&1 -\frac43\cos\theta\EXP{-l}+\frac13\EXP{-2b}+\frac13\EXP{-2l}
          -\frac43\cos\theta\EXP{-l-2b}+\EXP{-2l-2b} \label{detob1}\\
       &=&\frac43\EXP{-l-b}
          \left(
	    \sinh{b}\sinh{l}+2(\cosh{l}-\cos\theta)\cosh{b}
	  \right)
\:.\eea
Thus:
\be\label{Sob}
S(\gamma)=\sinh(\sqrt\gamma b)\sinh(\sqrt\gamma l)+
2[\cosh(\sqrt\gamma l)-\cos(\theta)]\cosh(\sqrt\gamma b)
\:,\ee
a result that can also be obtained using (\ref{MatriceM},\ref{MonPas}) (see
also appendix \ref{loops}).

\bigskip

As one can realize looking at (\ref{detob1}), there exists a symmetry between 
the coefficients of the different terms appearing in the diagrammatic 
expansion.
It is possible to use this symmetry to reduce the number of diagrams that 
have to be considered, which greatly simplify the calculation. This point
is discussed in more details in appendix \ref{SymmCoef}.

\bigskip

To conclude this section we insist on the fact that the periodic orbit 
expansion of the spectral determinant involves only a finite number of 
contributions, despite the number of primitive orbits is infinite. In 
contrast, the periodic orbit expansion of the partition function 
(\ref{RothIntro}) or the 
density of states involves an infinite number of contributions.


\section{Graphs in a magnetic field \label{Bfield}}

In this section, we describe the appropriate modifications to the previous 
formalism that have to be done in the presence of a magnetic field, 
namely for a distribution of Aharonov-Bohm fluxes. The operator of
interest is now $-(\D_x-\I A)^2$. In equation (\ref{CL2}) one has to
replace the derivative of the function $\psi$ by a covariant derivative
$\Dc_x=\D_x-\I A(x)$
\be\label{CL2wf}
\sum_\beta a_{\alpha\beta}\Dc_{x_{\alpha\beta}}
\psi_{(\alpha\beta)}({x_{\alpha\beta}=0})=0
\:.\ee
The path integral derivation of $\det(-\Dc_x^2+\gamma)$ follows
the same lines except that all derivatives have to be replaced by covariant
derivatives in (\ref{pi1},\ref{pi2}). One is led to an expression
of the determinant that involves a product of terms of the form
\be\label{contrib}
\int_{\phi(0)=\phi_\alpha}^{\phi(\lab)=\phi_\beta}
\dphi\EXP{-\frac12\int_0^{\lab}\D{x}\,(|\Dc_x\phi|^2+\gamma|\phi|^2)}
\:.\ee
We may recover the Gaussian action of the harmonic oscillator by
performing the following gauge transformation
\be
\phi(x)=\tilde\phi(x)\EXP{\I\int_{x_0}^{x}\D{x'}\,A(x')}
\ee
where the integral is performed along the bond $(\alpha\beta)$ and $x_0$ is
an arbitrary point on this bond (a change of $x_0$ corresponds to adding a
constant phase to the field). If we define 
\be
\theta_{\beta\alpha}=\int_0^{\lab}\D{x}\,A(x)
\ee
($\theta_{\beta\alpha}=-\theta_{\alpha\beta}$)
where the integral is performed along the bond $(\alpha\beta)$, and choose
$x_0$ in such a way that 
$\int_{x_0}^{\lab}\D{x}\,A(x)=\frac{\theta_{\beta\alpha}}{2}$ and
$\int_{x_0}^{0}\D{x}\,A(x)=\frac{\theta_{\alpha\beta}}{2}$, then
(\ref{contrib}) becomes
\be
\int_{\tilde\phi(0)   =\phi_\alpha\EXP{-\I\theta_{\alpha\beta}/2}}
    ^{\tilde\phi(\lab)=\phi_\beta \EXP{-\I\theta_{\beta\alpha}/2}}
{\cal D}\tilde\phi{\cal D}\bar{\tilde\phi}\:
\EXP{-\frac12\int_0^{\lab}\D{x}\,(|\D_x\tilde\phi|^2+\gamma|\tilde\phi|^2)}
=
\sqrt\gamma 
G_{\sqrt\gamma\lab}(\gamma^{1/4}\phi_\beta \EXP{-\I\theta_{\beta\alpha}/2},
                    \gamma^{1/4}\phi_\alpha\EXP{-\I\theta_{\alpha\beta}/2})
\:.\ee
Introducing this expression in (\ref{sp}), then it is clear that 
(\ref{MonPas}) still holds provided the matrix $M$ is now defined as
\be\label{MatriceM}
M_{\alpha\beta}=
\delta_{\alpha\beta}\sum_\mu a_{\alpha\mu}\coth(\sqrt{\gamma}l_{\alpha\mu})
-a_{\alpha\beta}
 \frac{\EXP{\I\theta_{\alpha\beta}}}{\sinh(\sqrt{\gamma}\lab)}
\:.\ee

The calculations of section \ref{Jean} may also be generalized. The
derivation is the same and the matrix $Q$ now reads
\be
Q_{(\alpha\beta)(\mu\nu)}=a_{\alpha\beta}a_{\mu\nu}
\delta_{\alpha\mu}\left(\frac{2}{m_\alpha}-\delta_{\beta\nu}\right)
\EXP{\I\frac{\theta_{\alpha\beta}-\theta_{\alpha\nu}}{2}}
\:,\ee
and the matrix $R$ is unchanged. To define $E$ it is convenient to introduce a
unitary transformation ${\cal U}$ that makes the changes more clear.
If one writes 
\be
Q R={\cal U}\,E^{\rm T}\,{\cal U}^\dagger
\:,\ee
where ${\cal U}$ is
\be
{\cal U}_{(\alpha\beta)(\mu\nu)}= a_{\alpha\beta}a_{\mu\nu}
\delta_{\alpha\mu}\delta_{\beta\nu}\EXP{\I\frac{\theta_{\alpha\beta}}{2}}
\:,\ee
then one has $E=D\epsilon$ where $\epsilon$ is still given by
(\ref{mateps}) and all the dependence in the fluxes is now contained in the
matrix $D$:
\be
D_{(\alpha\beta)(\mu\nu)}=a_{\alpha\beta}a_{\mu\nu}
\delta_{\alpha\mu}\delta_{\beta\nu}
\EXP{-\sqrt{\gamma}\,l_{\mu\nu}+\I\theta_{\mu\nu}}
\:.\ee
Since the spectral determinant is given by the determinant of the matrix
$1-E$, the matrix ${\cal U}$ and its inverse disappear in the determinant and
(\ref{resJean}) still holds. 

The derivation of the trace formulae only requires trivial modifications and
one finally obtains for the determinant:
\be\label{Sipwf}
S(\gamma)=\gamma^{\frac{V-B}{2}}\EXP{\sqrt{\gamma}\,L}
\Big(\prod_{\alpha}m_\alpha\Big)2^{-B}
\prod_{\tilde C}\left(
  1-\alpha(\tilde C)\EXP{-\sqrt{\gamma}\,l(\tilde C)+\I\theta(\tilde C)}\right)
\:,\ee
and the corresponding partition function:
\be
Z(t)=\frac{L}{2\sqrt{\pi t}}+\frac{V-B}{2}+\frac{1}{2\sqrt{\pi t}}
\sum_{C} l(\tilde C)\alpha(C)\EXP{-\frac{l(C)^2}{4t}+\I\theta(C)}
\ee
which generalizes the Roth's formula (\ref{Roth}). We have used the obvious 
notation for the flux enclosed by an orbit $C=(i_1,i_2,\cdots,i_n)$:
$\theta(C)=\theta_{i_1}+\cdots+\theta_{i_n}$.


\section{More general boundary conditions\label{MixBoun}}

One has sometimes to consider more general boundary conditions than
(\ref{CL2wf}). Imposing instead of (\ref{CL2wf}) the so-called mixed
boundary conditions
\be\label{CL2wfmb}
\sum_{\beta}a_{\alpha\beta}\Dc_{x_{\alpha\beta}}
\psi_{(\alpha\beta)}({x_{\alpha\beta}=0)}=\lambda_\alpha\psi_\alpha
\ee
requires to generalize some of the previous results\footnote{
One may distinguish two cases. 
({\it i}) In the Schr\"odinger problem, the current is 
$J=2\im(\psi^*\D_x\psi)$. Conditions (\ref{CL2wfmb}) also lead to current 
conservation at the vertices.
({\it ii}) For the diffusion problem, the wave function $\psi(x,t)$
has to be replaced by a probability density 
$P(x,t)$. The current of probability is given in this case by 
$J(x,t)=-\partial_xP(x,t)$. Equation (\ref{CL2wfmb}) implies that such a 
current is not conserved at the vertices where $\lambda_\alpha\neq0$. This 
condition corresponds physically to a graph connected to the external world, 
$\lambda_\alpha P(\alpha,t)$ being the 
current of particles exiting the graph. In the Dirichlet case 
($\lambda_\alpha=\infty$) the graph is perfectly connected, which means that 
a particle exits the graph with probability $1$ if it reaches the vertex
$\alpha$.
}. This can be easily
achieved by using the path integral formalism. It is rather obvious
that the boundary terms in (\ref{pi2}) now produce additional Gaussian term in
(\ref{sp}) which now reads 
\be
S(\gamma)^{-1}=\gamma^{\frac{B-V}{2}}
\int \prod_{\alpha=1}^V\D\phi_\alpha\D\bar\phi_\alpha\,
\EXP{-\frac{\lambda_\alpha}{2\sqrt\gamma}|\phi_\alpha|^2}
\prod_{(\alpha\beta)}
G_{\sqrt{\gamma}\lab}(\phi_\beta \EXP{-\I\theta_{\beta\alpha}/2},
                      \phi_\alpha\EXP{-\I\theta_{\alpha\beta}/2})
\:.\ee
Thus only the diagonal elements of matrix $M$ change. In (\ref{MonPas}), 
$M$ is replaced by $\tilde M$:
\be\label{Mtil}
\tilde{M}_{\alpha\beta}=M_{\alpha\beta}
                       +\frac{\lambda_\alpha}{\sqrt\gamma}\delta_{\alpha\beta}
\:.\ee

It is also easy to see from the discussion of section \ref{Jean} how 
the different matrices are affected by the change of boundary conditions. 
Equation (\ref{SJ1}) will receive some additional Gaussian terms 
$\EXP{-\frac12\lambda_\alpha|\phi_\alpha|^2/\sqrt\gamma}$ which implies that
$m_\alpha$ has to be replaced by $m_\alpha+\frac{\lambda_\alpha}{\sqrt\gamma}$.
The scattering matrix $Q$ now reads
\be
Q_{(\alpha\beta)(\mu\nu)}=a_{\alpha\beta}a_{\mu\nu}
\delta_{\alpha\mu}
\bigg(
  \frac{2}{m_\alpha+\frac{\lambda_\alpha}{\sqrt\gamma}}-\delta_{\beta\nu}
\bigg)\:\EXP{\I\frac{\theta_{\alpha\beta}-\theta_{\alpha\nu}}{2}}
\ee
and the matrix $\epsilon$ is
\be\label{epsmb}
\epsilon_{(\alpha\beta)(\mu\nu)}=a_{\alpha\beta}a_{\mu\nu}
\delta_{\beta\mu}
\bigg(
  \frac{2}{m_\beta+\frac{\lambda_\beta}{\sqrt\gamma}}-\delta_{\alpha\nu}
\bigg)
\:,\ee
the other matrices $R$ and $D$ remain unchanged. The factor $\alpha(C)$ of
sections \ref{TraFor} and \ref{DiagExp} will be modified according to
(\ref{epsmb}). 
 
\bigskip

Expression (\ref{Mtil}) allows us to treat both the case of Neumann boundary 
conditions ($\lambda_\alpha=0$) which has already been studied in this paper, 
and the case of Dirichlet boundary condition ($\lambda_\alpha=\infty$).
For a diffusive conductor, these conditions correspond to a disconnected wire
and a wire perfectly connected to leads at the node, respectively.

If one imposes Dirichlet condition at all vertices of the graph, one has
to consider the limit $\lambda\to\infty$ for all vertices. Then 
$\det(\tilde{M})\simeq\prod_\alpha\frac{\lambda_\alpha}{\sqrt\gamma}$. 
The spectral determinant is 
$S(\gamma)\simeq\Big(\prod_\alpha\lambda_\alpha\Big)
\prod_{(\alpha\beta)}\frac{\sinh\sqrt\gamma\lab}{\sqrt\gamma}$. 
One can drop the irrelevant factor $\prod_\alpha\lambda_\alpha$ since 
the spectral determinant is defined up to a multiplicative numerical factor 
independent of $\gamma$, that depends on the regularization.
One finds that the spectral determinant 
$S(\gamma)=\prod_{(\alpha\beta)}\frac{\sinh\sqrt\gamma\lab}{\sqrt\gamma}$ is
the product of the spectral determinants associated with each bonds. All bonds
are then independent.

We now discuss the case where one imposes Dirichlet 
boundary at only one vertex $\alpha_0$ which will be useful in the next 
section. Taking the limit $\lambda_{\alpha_0}\to\infty$, one has:
\be\label{Diraop}
S(\gamma;\{{\rm Dirichlet\:at\:}\alpha_0\})
=\gamma^{\frac{V-B-1}{2}} \
\prod _{(\alpha \beta)} \sinh(\sqrt{\gamma }l_{\alpha\beta}) 
\:\det(M^{\alpha_0})
\ee
where $M^{\alpha_0}$ is the $(V-1)\times(V-1)$ matrix given by the matrix $M$ 
with the line $\alpha_0$ and the column $\alpha_0$ deleted.

\bigskip

As an example consider the graph pictured in figure \ref{ouvrebouteille} for
which one imposes Dirichlet boundary at the end of the arm. $S(\gamma)$ 
becomes:
\be\label{SobwD}
S_{\stackrel{\rm Dirich.\:at}{\rm end\:of\:arm}}(\gamma)=
\frac{1}{\sqrt\gamma}\Big\{
  \cosh(\sqrt\gamma b)\sinh(\sqrt\gamma l)+
  2[\cosh(\sqrt\gamma l)-\cos(\theta)]\sinh(\sqrt\gamma b)
\Big\}
\:,\ee
to be compared with (\ref{Sob}).
It is interesting to check that the magnetization of the ring, given by
(\ref{currentee3},\ref{Mtyp}), does not depend on the
choice of the boundary conditions at the end of the arm, when its length
$b$ goes to infinity ($b\gg L_\phi=1/\sqrt\gamma$).


\section{Zero mode and low energy behaviour\label{ZeroMode}}

The low energy part of the spectrum can be studied by using the expansion of
the spectral determinant when $\gamma\to0$. Using this approach we prove the 
existence of a zero mode state when ${\cal B}=0$ and obtain the dependence of 
the ground state energy in the magnetic field. Using the infinite product 
representation $S(\gamma)=\prod_n(E_n+\gamma)$ and assuming $E_0=0$
implies that $S(\gamma)$ behaves linearly with $\gamma$ when $\gamma\to0$.
In order to study this limit it is convenient to expand $M$ as a power series
in $\gamma$
\be
M(\gamma)=\frac{1}{\sqrt\gamma}\hat{M}(\gamma)
         =\frac{1}{\sqrt\gamma}(\hat{M}^0 + \gamma\hat{M}^1+\cdots )
\ee
where 
\bea
\hat{M}^0_{\alpha\beta} &=& \delta_{\alpha\beta}\sum_\mu
\frac{a_{\alpha\mu}}{l_{\alpha\mu}} - \frac{a_{\alpha\beta}}{l_{\alpha\beta}}\\
\hat{M}^1_{\alpha\beta} &=& \delta_{\alpha\beta}\frac13\sum_\mu
{a_{\alpha\mu}}{l_{\alpha\mu}} + \frac16{a_{\alpha\beta}}{l_{\alpha\beta}}
\:.\eea
From these expressions, it is easy to see that $\hat{M}^0$ possesses an  
eigenvalue $\xi_0=0$ corresponding to the eigenvector $v_0$ whose 
components  are $(v_0)_\alpha=1$. 
Remembering that $M$ acts on the $V$-vector constructed with the wave 
function at the nodes, the vector $v_0$ corresponds to a wave function that 
takes the same value on all vertices and is constant on the graph 
(since $\gamma=-E=0$). 
We denote by $\xi_n(\gamma)$ ($n=0,\cdots,V-1$) the $V$ eigenvalues of 
$\hat{M}(\gamma)$. The eigenvalue $\xi_0(\gamma)$ vanishes at $\gamma=0$ 
and its behaviour for small $\gamma$ computed in perturbation theory is
$\xi_0(\gamma)\simeq\gamma\frac{v_0^{\rm T}\hat{M}^1v_0}{v_0^{\rm T}v_0}
=\frac{\gamma L}{V}$. 
This implies that
$
\det(\hat{M}(\gamma))\simeq\frac{\gamma L}{V}\prod_{n=1}^{V-1}\xi_n(0)
$.
It is possible to write the product of the non-vanishing eigenvalues for 
$\gamma=0$ as the determinant of the matrix
$\hat{M}^0+\frac{v_0\,v_0^{\rm T}}{||v_0||^2}$ where the second term is the 
projector on the vector $v_0$.

One finally obtains for the spectral determinant 
\be\label{Sapg}
S(\gamma)\APPROX{\gamma\to0}\gamma
\frac{L}{V}\Big(\prod_{(\alpha\beta)}l_{\alpha\beta}\Big)\det(K)
\ee
where the matrix $K$ is 
\be
K_{\alpha\beta}=\hat{M}^0_{\alpha\beta}+\frac1V
\:.\ee

From the linear behaviour $S(\gamma)\propto\gamma$ at small $\gamma$ it 
follows that one has for the partition function: 
$\lim_{t\to\infty}Z(t)=1$. 

Let us remark that if all the lengths of the bonds are equal to unity, the 
product $\prod_{n=1}^{V-1}\xi_n(0)$ is equal to ${\cal T}({\cal G})$, 
the number of trees that cover the graph 
(Tutte theorem) \cite{Big76,CveDooSac80}.

\bigskip

It is also interesting to see how the spectral determinant behaves at non 
vanishing magnetic field. For the sake of simplicity we denote the ensemble
of fluxes $\{\theta_{\alpha\beta}\}$ by ${\cal B}$. The matrix $\hat{M}$ now
depends on the fluxes. The spectral determinant
may be expressed in terms of the eigenvalues of $\hat{M}(\gamma,{\cal B})$:
\be
S(\gamma,{\cal B})=
\prod_{(\alpha\beta)}\frac{\sinh\sqrt\gamma\lab}{\sqrt\gamma}
\prod_{n=0}^{V-1}\xi_n(\gamma,{\cal B})
\:.\ee
At zero magnetic field and small $\gamma$, one has shown above how 
$S(\gamma)$ behaves (\ref{Sapg}) by a perturbative expansion of 
$\xi_0(\gamma,0)$. Similarly, at small magnetic field, one may compute 
the eigenvalue $\xi_0(0,{\cal B})$ in perturbation theory. Starting from
\be
\hat{M}_{\alpha\beta}(0,{\cal B})=
\delta_{\alpha\beta}\sum_\mu\frac{a_{\alpha\mu}}{l_{\alpha\mu}} 
- a_{\alpha\beta}\frac{\EXP{\I\theta_{\alpha\beta}}}{\lab}
\ee
and expanding this expression at small fluxes, it is easy to show that 
$\xi_0(0,{\cal B})\simeq
\frac{1}{2V}\sum_{\alpha,\beta}a_{\alpha\beta}
\frac{\theta_{\alpha\beta}^2}{\lab}$.
Since $\prod_{n=1}^{V-1}\xi_n(0,0)=\det(K)$ the spectral determinant 
behaves like
\be
S(0,\left\{\theta_{\alpha\beta}\right\})
\APPROX{\{\theta_{\alpha\beta}\}\to0}
\frac1V
\sum_{(\alpha\beta)}\frac{\theta_{\alpha\beta}^2}{\lab}
\Big(\prod_{(\alpha\beta)}\lab\Big)\det(K)
\:.\ee

One may extract from these expressions the ground state energy at small 
magnetic field. At small $\gamma$ one has: 
$S(\gamma,0)\simeq\gamma\prod_{n>0}E_n(0)$. On the other hand, when the 
magnetic field goes to zero, the ground state energy is the only energy to
vanish, then $S(0,{\cal B})\simeq E_0({\cal B})\prod_{n>0}E_n(0)$. The ratio 
$\frac{S(0,{\cal B})}{S(\gamma,0)}\simeq\frac{E_0({\cal B})}{\gamma}$ gives
the expression of the ground state energy
\be\label{gs}
E_0\left(\left\{\theta_{\alpha\beta}\right\}\right)
\simeq\frac{1}{L}\sum_{(\alpha\beta)}\frac{\theta_{\alpha\beta}^2}{\lab}
\:.\ee

Let us notice that the expression (\ref{gs}) 
of the ground state energy at small magnetic field may be recovered in a 
simpler way by perturbation theory. The eigenfunction of the ground state 
at zero field is $\psi_0(x\in{\cal G})=\frac{1}{\sqrt{L}}$. Thus the 
correction to the energy at small field is 
$
E_0(\{\theta_{\alpha\beta}\})\simeq \bra{\psi_0}A(x)^2\ket{\psi_0}
=\frac{1}{L}\sum_{(\alpha\beta)}\int_0^{\lab}\D{x}\,A(x)^2
$.
By choosing a gauge such that $A(x)=A_{\beta\alpha}$ is constant
on the bond $(\alpha\beta)$, one recovers (\ref{gs}) since 
$\theta_{\beta\alpha}=A_{\beta\alpha}\lab$.

\bigskip

So far we did not describe the effect of a change of boundary conditions
on the low energy behaviour of the spectral determinant. 
The matrix $M$ at zero magnetic field now becomes
\be
\hat{\tilde M}_{\alpha\beta}(0,\{\lambda_\alpha\})=
\hat{M}^0_{\alpha\beta}+\lambda_\alpha\delta_{\alpha\beta}
\:.\ee
As in the presence of a small magnetic field one may compute the lowest
eigenvalue $\xi_0(0,\{\lambda_\alpha\})$ of the matrix 
$\hat{\tilde M}(0,\{\lambda_\alpha\})$ by perturbation theory. This gives
\be
S(0,\{\lambda_\alpha\})
\APPROX{\{\lambda_\alpha\}\to0}
\frac1V\sum_{\alpha}\lambda_\alpha
\Big(\prod_{(\alpha\beta)}\lab\Big)\det(K)
\:.\ee
One may also deduce from this result the behaviour of the ground state
energy:
\be\label{gsmb}
E_0\left(\{\lambda_\alpha\}\right)
\simeq\frac{1}{L}\sum_{\alpha}\lambda_\alpha
\:.\ee

Let us notice that in a particular case, one may recover this result by 
another way. Consider a ring with $V$ vertices with mixed boundary 
conditions. This problem is equivalent to a ring with a potential made of
$\delta$ scatterers: $W(x)=\sum_{\alpha}\lambda_\alpha\delta(x-x_\alpha)$, 
where the $x_\alpha$'s are the positions of the vertices on the ring. In this 
case it is possible to compute perturbatively the energy of the ground state:
$E_0\left(\{\lambda_\alpha\}\right)\simeq\bra{\psi_0}W(x)\ket{\psi_0}$ 
which indeed leads to (\ref{gsmb}).

The result (\ref{gsmb}) has a physical interpretation in the context of 
 diffusion. The zero mode
$E_0=0$ is associated with the uniform stationary distribution at long time.
As soon as the graph is connected to the external world, which 
may be described by 
taking mixed boundary conditions, the probability to remain on the graph
decreases at long time $\int_{\rm Graph}\D x\,P(x,x',t)\sim\EXP{-E_0 t}$ 
which means absence of a zero mode. In this case, $E_0$ is the inverse
escape time of the graph.


\section{The graph connected to an infinite lead \label{Scatt}}

In this section, we consider the case of a graph connected to an infinite 
lead attached to the vertex $\alpha$. In that case, we are dealing with a 
quantum
scattering problem where the relevant information like the spectrum, the 
Wigner time delays or thermodynamic quantities like the persistent currents 
are encoded in the (unitary) scattering matrix \cite{AkkAueAvrSha91}. 
To the purpose of calculating the scattering phase shifts, let us suppose that
along the lead an incoming plane wave $\EXP{-\I kx}$ enters the graph
at the vertex $\alpha$ and is reflected with a phase shift $\delta_\alpha(E)$.
Below we derive a formula expressing the phase shift in terms of $\det(M)$. 
Although such a formula is in fact a special case of the one given in 
\cite{KotSmi99,KotSmi99a}, it is nevertheless interesting to present an 
independent derivation of it and show the relation with spectral determinants.
\begin{figure}[!h]
\begin{center}
\includegraphics{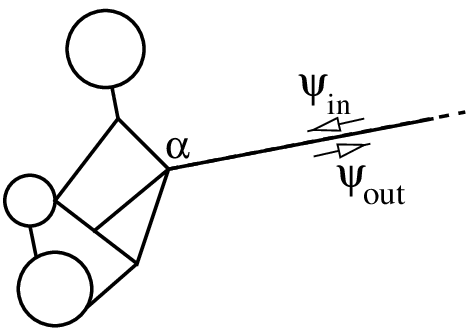}
\caption{}
\end{center}
\end{figure}
$m_\alpha$ is the coordinence of vertex $\alpha$ in the 
absence of the external lead. We first write the wave function of energy 
$E=k^2$ on the bond $(\mu\beta)$ in a form that insures continuity:
\be
\psi_{(\mu\beta)}(x_{\mu\beta})
=\frac{ \EXP{\I A_{\beta\mu}x_{\mu\beta}} }{ \sin kl_{\mu\beta} }
\left(
   \psi_\mu \sin k(l_{\mu\beta}-x_{\mu\beta})
  +\psi_\beta\EXP{-\I A_{\beta\mu}l_{\mu\beta}}\sin k x_{\mu\beta}
\right)
\ee
where $A_{\beta\mu}l_{\mu\beta}=\theta_{\beta\mu}$.
On the incoming lead, the wave function is written in terms of 
stationary scattering states
\be
\psi_{\rm lead}(x)=\frac{\psi_\alpha}{\cos(\delta_\alpha/2)}
\cos(kx+\delta_\alpha/2)
\propto\EXP{-\I kx} + \EXP{\I kx +\I\delta_\alpha}
\ee
where $x\in[0,\infty[$ is a current point on the lead.
Continuity of the wave function at vertex $\alpha$ is already insured with 
the previous expression.
Current conservation at the vertex $\alpha$ gives
\be
\sum_\beta a_{\alpha\beta}\left(
  -\psi_\alpha \cotg k\lab 
  +\psi_\beta \frac{\EXP{\I\theta_{\alpha\beta}}}{\sin k\lab}
\right)
- \psi_\alpha \tan(\delta_\alpha/2) = 0
\:.\ee
Since the boundary conditions on the other vertices are obviously unchanged
we are left with the linear system of $V$ equations:
\be
-\I\sum_\beta M_{\mu\beta}(\gamma=-k^2) \psi_\beta
= \delta_{\mu\alpha} \psi_\alpha \tan(\delta_\alpha/2)
\ee
where $\mu=1,...,V$.
Cramer's formula then gives
\be
\psi_\alpha = 
\I \tan(\delta_\alpha/2) \frac{\det(M^{\alpha}(-k^2))}{\det(M(-k^2))} 
\psi_\alpha 
\ee
where $M^{\alpha}(\gamma)$ is the matrix introduced in section \ref{MixBoun}.
Then the phase shift is obviously given by
\be
\cotg(\delta_\alpha(E)/2)=\I\frac{\det(M^{\alpha}(-E))}{\det(M(-E))}
\:.\ee
This result, together with (\ref{Diraop}), shows 
that the phase shift is related to the ratio of two spectral determinants: one 
calculated with a Dirichlet boundary at the vertex connected to the lead and
Neumann boundary at all other vertices (\ref{Diraop}), and the other 
calculated with Neumann boundary at all vertices (\ref{detpasmon}):
\be
\cotg(\delta_\alpha(k^2)/2)
=-k\frac{S(-k^2;\{{\rm Dirichlet\:at\:}\alpha\})}{S(-k^2)}
\:.\ee
This expression shows that $\delta_\alpha=0\:{\rm mod}\:2\pi$ if $k^2$ 
coincides with an energy of the graph with Neumann boundary, and 
$\delta_\alpha=\pi\:{\rm mod}\:2\pi$ if $k^2$ is an energy of the 
graph with Dirichlet at vertex $\alpha$.

\bigskip

As an example the phase shift of the graph of figure \ref{ouvrebouteille} is
given by the ratio of (\ref{SobwD}) and (\ref{Sob}) if the lead is connected 
to the end of the arm
\be
\cotg(\delta/2)=
\frac{\cos kb \sin kl + 2(\cos kl - \cos\theta) \sin kb}
     {\sin kb \sin kl - 2(\cos kl - \cos\theta) \cos kb}
\:.\ee

As a second example the phase shift in the case of the complete graph $K_n$ 
that will be studied in the next section reads
\be
\cotg(\delta/2)
=\cos(\varphi)\frac{\cos(k\ell)+\cos(\varphi)-1}{\cos(k\ell)+\cos(\varphi)}
\cotg(k\ell/2)
\:,\ee
where $\cos(\varphi)=\frac{1}{n-1}$. As expected the scattering length 
$\frac{\D\delta}{\D k}\big|_{k=0}=2\frac{n(n-1)}{2}\ell=2L$ is a measure of 
the total length of the graph.


\section{Application to the case of the complete graph $K_n$ \label{CompGra}}

\begin{figure}[!h]
\begin{center}
\includegraphics{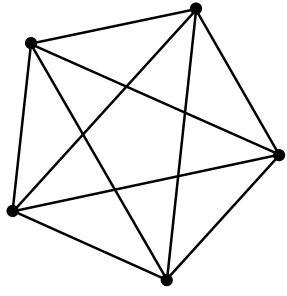}
\end{center}
\caption{The complete graph $K_5$.}
\label{k5}
\end{figure}
As an illustration of the previous formalism it is interesting to consider the 
particular case where the graph $\cal{G}$ is the complete graph $K_n$ whose 
$V=n$ vertices are all connected (see figure \ref{k5}). Then the number of 
bonds is $B=\frac{n(n-1)}{2}$.  

In the case where all the bonds of a graph have the same length $\ell$ one 
obtains 
\be
\det(M)=\left(\frac{1}{\sinh\sqrt\gamma\ell}\right)^n\: 
P((n-1)\cosh(\sqrt\gamma\ell))
\ee
where 
$P(X)=\det(X\delta_{\alpha\beta}-a_{\alpha\beta})$ is the characteristic
polynomial of the graph.
The characteristic polynomial of $K_n$ can be found in classical textbooks.
For the complete graphs $a_{\alpha\beta}=1-\delta_{\alpha\beta}$,
thus the determinant involved in $P(X)$ is of the same form as the 
determinant of matrix $F$ given in appendix \ref{MatProp}. This gives 
the following expression for the spectral determinant
\be
S(\gamma)=
\left({\sqrt\gamma}\right)^{\frac{3n}{2}-\frac{n^2}{2}}
\left(\sinh\sqrt\gamma\ell\right)^{\frac{n^2}{2}-\frac{3n}{2}} (n-1)
(\cosh\sqrt\gamma\ell -1)\left[(n-1)\cosh\sqrt\gamma\ell +1\right]^{n-1}
\:.\ee
By calculating the inverse Laplace transform of 
$\drond{}{\gamma}\ln{S(\gamma)}$ one obtains the partition function
\bea\label{ZKnRoth}
Z(t)&=&\frac{n(n-1)\ell}{4\sqrt{\pi t}} + \frac{n(3-n)}{4} +
\frac{\ell}{2\sqrt{\pi t}}
\bigg\{
  \left[ \theta\left(\frac{\ell^2}{4\pi t}\right) - 1 \right]  
  \nonumber \\ 
  & & \hspace{1.5cm}
  + \frac{n(n-3)}{2}\left[ \theta\left(\frac{\ell^2}{\pi t}\right) - 1 \right]
  + 2(n-1)\sum_{k=1}^\infty (-1)^k T_k\left(\frac{1}{n-1}\right)
                            \EXP{-\frac{k^2\ell^2}{4t}}
\bigg\}
\eea
where $\theta(x)=\sum_{k=-\infty}^{+\infty}\exp{-\pi k^2x}$ is the Jacobi 
$\theta$ function, and $T_k(x)$ Tchebychev polynomials.

\subsection*{a) short time limit} 

\begin{figure}[!h]
\begin{center}
\includegraphics{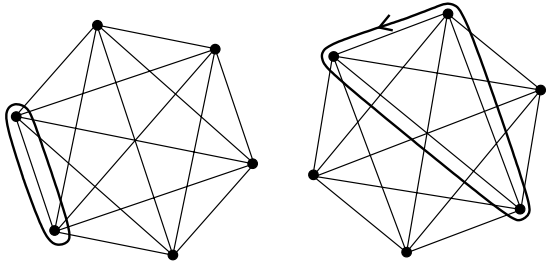}
\end{center}
\caption{Some orbits associated with the first terms in the brackets of 
         (\ref{KnRexp}) for the complete graph $K_6$.}
\label{k6}
\end{figure}

Equation (\ref{ZKnRoth}) is written in the form of Roth formula, appropriate 
for short times. The first contributions coming from the expansion
of (\ref{ZKnRoth}) in the limit $t\to0$ give
\bea\label{KnRexp}
Z(t)&=&\frac{L}{2\sqrt{\pi t}} + \frac{V-B}{2} + \frac{1}{2\sqrt{\pi t}}
\bigg\{
  \frac{n(n-1)}{2} 2\ell \left(\frac{2}{n-1}-1\right)^2
  \EXP{-\frac{\ell^2}{t}}
  \nonumber\\ & & \hspace{4cm}
  + 2\frac{n(n-1)(n-2)}{3!} 3\ell \left(\frac{2}{n-1}\right)^3
  \EXP{-\frac{9\ell^2}{4t}}+\cdots
\bigg\}
\eea
The first two terms coincide with the first terms of Roth formula. For 
the two following terms we recognize the contributions of the shortest orbits 
represented in figure \ref{k6}. For the orbits of length $2\ell$, the factor 
$\left(\frac{2}{n-1}-1\right)^2$ is the weight $\alpha(C)$ of those orbits 
and $\frac{n(n-1)}{2}$ their number. The next term corresponds to orbits
of length $3\ell$ with weight $\alpha(C)=\left(\frac{2}{n-1}\right)^3$. Their
number is $2\frac{n(n-1)(n-2)}{3!}$ where the additional factor $2$ comes
from time reversed orbits.

\subsection*{b) Long time limit} 

For the graph $K_n$ it is also possible to study the limit $t\to\infty$.
Introducing $\varphi$, defined by $\cos\varphi=\frac{1}{n-1}$, one may use
the identity $T_k(\cos\varphi)=\cos{k\varphi}$ and the Poisson summation
formula to find
\be\label{ZKn}
Z(t)=\theta\left(\frac{4\pi t}{\ell^2}\right) +
\frac{n(n-3)}{4}
\left[\theta\left(\frac{\pi t}{\ell^2}\right)-1\right]
+(n-1)\sum_{k=-\infty}^{+\infty}
\EXP{-\frac{4\pi^2}{\ell^2}[k-\frac{1}{2\pi}(\varphi+\pi)]^2 t}
\:.\ee
It is easy to see that 
\be
Z(t)=1+O\left(\EXP{-\kappa t}\right)
\ee
where the contribution of the first term comes from the constant zero mode 
$\psi_0(x\in{\cal G})=\frac{1}{\sqrt{L}}$.

\subsection*{c) The eigenvalue spectrum} 

Equation
(\ref{ZKn}) provides the whole spectrum of $K_n$. The first two terms describe
a series of states $E_{12,k}=k^2\frac{\pi^2}{\ell^2}$, $k\in\NN$ with 
degeneracies $d_{12,k=0}=1$, $d_{12,k}=\frac{n(n-3)}{2}$ if $k$ is odd and 
$d_{12,k}=2+\frac{n(n-3)}{2}$ if $k$ is even. The third term generates 
energies $E_{3,k}=\frac{\pi^2}{\ell^2}\left(2k-1-\frac{\varphi}{\pi}\right)^2$,
$k\in\ZZ$ with degeneracies $d_{3,k}=n-1$. These states are similar to those 
of a ring pierced by a flux, where the parameter $\varphi$ related to the 
coordinence plays the role of the flux. 

For $n\geq3$, $\frac{\varphi}{\pi}\in[\frac13,\frac12[$ and one may 
reorganize the energy levels:

\medskip

\renewcommand{\arraystretch}{1.25}

\begin{center}
\begin{tabular}{ll}
Energy                 & Degeneracy             \\
\hline\hline
$E_0=E_{12,0}=0$       & $d_0=1$                \\
$E_1=E_{3,1}=\frac{\pi^2}{\ell^2}\left(1-\frac{\varphi}{\pi}\right)^2$
                       & $d_1=n-1$              \\
$E_2=E_{12,1}=\frac{\pi^2}{\ell^2}$
                       & $d_2=\frac{n(n-3)}{2}$ \\
$E_3=E_{3,0}=\frac{\pi^2}{\ell^2}\left(1+\frac{\varphi}{\pi}\right)^2$
                       & $d_3=n-1$              \\
$E_4=E_{12,2}=\frac{\pi^2}{\ell^2}2^2$
                       & $d_4=2+\frac{n(n-3)}{2}$ \\
$E_5=E_{3,2}=\frac{\pi^2}{\ell^2}\left(3-\frac{\varphi}{\pi}\right)^2$
                       & $d_5=n-1$              \\
$\ \ \ \ \ \ \vdots$   & $\ \ \ \vdots$         \\
\hline
\end{tabular}
\end{center}

\renewcommand{\arraystretch}{1}

\medskip

\noindent
The case $n=3$ corresponds to the ring of length $3\ell$. In this case 
$\varphi/\pi=1/3$ and one recovers the well-known spectrum.

It is interesting to note that despite the particle explores a volume 
$L\simeq{\frac12n^2\ell}$ at large $n$, the energy of the first excited state 
$E_1\simeq\frac{\pi^2}{4\ell^2}$ is not of order $\frac{1}{L^2}$, 
as one could have naively guessed, but instead of order $\frac{1}{\ell^2}$.


\section{Conclusion}

We have investigated spectral properties of the Laplacian on graphs by 
providing several equivalent representations of the corresponding spectral 
determinant. This has been achieved thanks to a path integral formulation 
and to the fact that the spectral determinant can be written in terms of the 
propagator of a 2D harmonic oscillator thus leading to a set of 
straightforward Gaussian integrals. We have thus obtained an expression of 
the spectral determinant as a trace formula involving the contribution of an 
infinite number of periodic orbits. Using a systematic diagrammatic method, 
we expressed the finite number of terms of the determinant as contributions 
of a finite number of orbits. Although 
it has been already clear in the literature that such a reduction is possible, 
the present formalism allows us to implement it directly for any given graph. 

The flexibility and the relative simplicity of this formalism is a hint for 
using it in a broader range of problems. For networks of mesoscopic and 
coherent conductors, it may help to compute both thermodynamic and transport 
properties in some non trivial situations such as the local distribution of 
persistent currents in a network driven by a far remote Aharonov-Bohm flux. 
Along the same lines, it could be possible to address the problem of 
topological properties of fractal networks like the Sierpinski gasket 
subject to a distribution of Aharonov-Bohm fluxes. 


\section*{Acknowledgments}

One of us (C.T.) was partially supported for this work by the Swiss National 
Science Foundation and by the TMR Network Dynamics of Nanostructures. 


\begin{appendix}

\section{The spectral determinant obtained by constructing the Green's
         function on the graph\label{DemGilles}}

We want to solve the diffusion equation
\be\label{G1}
(\gamma-\D^2_x) G(x,y)=\delta(x-y)
\ee
on a graph made of $V$ vertices (or nodes) linked by $B$ bonds. $y$ is the 
source for the diffusion. The solution of (\ref{G1}) is of the  form:
\be
G(x,y)= \sum_n {\psi_n(x) \psi_n^*(y) \over \gamma+E_n}
\ee
where $\psi_n$ are the eigenfunctions of the operator $-\D_x^2$. The spectral 
determinant is obtained by spatial integration of the diagonal Green's 
function  $G(y,y)$ :
\be 
\int\D{y}\, G(y,y) = \sum_n {1 \over \gamma + E_n} 
= {\partial \over \partial \gamma} \ln S(\gamma)
\:.\ee 

For a given source located in $y$, the diffusion equation is solved on each 
bond $(\alpha\beta)$ in terms of the values $G(\alpha,y)$ and $G(\beta,y)$ at 
the nodes:
\be\label{G3}
G(x,y)=G(\alpha,y) \cosh \sqrt{\gamma} x + \left( G(\beta,y)-G(\alpha,y) 
\cosh \sqrt{\gamma} l_{\alpha\beta} \right) 
{\sinh \sqrt{\gamma} x \over \sinh \sqrt{\gamma} l_{\alpha\beta}}
\ee
where $x$ is the linear coordinate on the bond $(\alpha\beta)$ of length 
$l_{\alpha\beta}$.

Current conservation at the vertex $\alpha$:
\be\label{G2}
- \sum_{\beta} \D_{x_{\alpha\beta}} G(x_{\alpha\beta}=0,y)=\delta_{\alpha,y}
\:,\ee
where the sum stands over the nearest vertices of $\alpha$, leads to the 
following equations
\be\label{G4}
G(\alpha,y) \sum_\beta \coth \eta_{\alpha\beta} 
          - \sum_\beta {G(\beta,y) \over \sinh \eta_{\alpha\beta}}
={\delta_{\alpha,y} \over \sqrt{\gamma}}
\ee
where $\eta_{\alpha\beta}=\sqrt{\gamma} l_{\alpha\beta}$. We have thus 
obtained a system of $(V+1)$ linear equations for the $(V+1)$ variables 
$G(\alpha, y)$ where $\alpha$ can be either a node of the graph or the source 
point $y$.
\begin{equation} M_y 
\left(
\begin{array}{c}
G(\alpha_{1},y) \\ \vdots \\ G(y,y) \\ \vdots \\ G(\alpha_{V},y)
\end{array}
\right)
=
\left(
\begin{array}{c}
0 \\ \vdots \\ 1 / \sqrt{\gamma} \\ \vdots \\ 0
\end{array}
\right)
\end{equation}
where the $(V+1)\times(V+1)$ matrix $M_y$ is defined by equations 
(\ref{M1},\ref{M2}). 
Here, $\alpha$ and $\beta$ are either vertices of the graph or the source 
point located in $y$. One now wants to calculate
$G(y,y)$. First, using (\ref{G4}), it is written in terms of $G(a,y)$ and 
$G(b,y)$ where $a$ and $b$ are the vertices ending the bond to which $y$ 
belongs:
\be\label{G5}
G(y,y) \left( \coth \eta_{a y}+\coth \eta_{y b} \right) -
{G(a,y) \over \sinh \eta_{a y}} - {G(b,y) \over \sinh \eta_{b y}}   
={1 \over \sqrt{\gamma}}
\ee
so that:
\be\label{G9}
G(y,y)={1 \over \sinh   \eta_{a b}} 
\left( {\sinh \eta_{a y} \sinh \eta_{b y} \over   
\sqrt{\gamma}}+G(a,y)  \sinh \eta_{b y} + G(b,y) \sinh \eta_{a y} \right)
\:.\ee

\bigskip

The variable $G(y,y)$ can then be eliminated in the previous system. The two 
equations for $G(a,y)$ and $G(b,y)$ are modified as:
\begin{equation}
G(a,y) \sum_{\beta} 
\coth \eta_{a\beta}
- \sum_{\beta} G(\beta,y) 
{1 \over \sinh \eta_{a \beta}}
= {1 \over \sqrt{\gamma} } 
\frac {\sinh \eta_{y b}}
{\sinh \eta_{ab}}
\end{equation}
and the $(V-2)$ other equations are unchanged. We obtain now a system for the 
$V$ variables $G(\alpha,y)$: 
\begin{equation}
M\:\left(
\begin{array}{c}
G(\alpha_{1},y) \\ \vdots \\ G(a,y) \\ \vdots \\
G(b,y) \\ \vdots \\ G(\alpha_{V-2},y)
\end{array}
\right)
={1 \over \sqrt{\gamma} \sinh \eta_{ab} }
\left(
\begin{array}{c}
0 \\ \vdots \\ 
\sinh \eta_{yb} 
\\ \vdots \\
\sinh \eta_{ay}
\\ \vdots \\ 0
\end{array}
\right)
\end{equation}
where $M$ is defined in (\ref{M1},\ref{M2}). $a$ and $b$ are now two of 
the $V$ vertices and the source point $y$ is now excluded. 

By inversion of the matrix $M$, one obtains $G(a,y)$ and $G(b,y)$:
\be
G(a,y)  =  {1 \over \sqrt{\gamma} \sinh \eta_{ab} }
\left( T_{aa}\sinh \eta_{yb} + T_{ab}\sinh \eta_{ay}\right)
\ee
where $T=M^{-1}$. From (\ref{G9}), one finally obtains $G(y,y)$:
\be
G(y,y) = \frac{1}{\sqrt{\gamma}}
\left \lbrace
\frac
{\sinh \eta_{ay} 
\sinh \eta_{yb} }
{\sinh \eta_{ab}}
+ \frac {1}{\sinh^2 \eta_{ab}}
\left[
T_{aa} \sinh^2 \eta_{by} +T_{bb} \sinh^2 \eta_{ay}+ 2 T_{ba}
\sinh \eta_{ay} 
\sinh \eta_{yb} 
\right]
\right \rbrace
\:.\ee
The  spatial integral of $G(y,y)$ on the bond $(ab)$ can be written as:
\bea
\int_a^b\D{y}\, G(y,y) &=& \frac{1}{2 \gamma } 
\Big\{ 
\eta_{ab} \coth  \eta_{ab}-1 
\nonumber \\
& & \hspace{0.5cm} + 
(T_{aa}+T_{bb})(1+2 \gamma 
{\partial \over \partial \gamma}) \coth \eta_{ab}- 2 T_{ba} 
(1+2 \gamma {\partial \over \partial \gamma})
{1 \over \sinh \eta_{ab}} \Big\}
\eea
where we have used the equalities:
\begin{eqnarray}
{-\eta \over \sinh^2 \eta}
&=& 2 \gamma {\partial \over \partial \gamma} \coth \eta \\ 
-\eta{\cosh \eta \over \sinh^2 \eta}\:,
&=& 2 \gamma{\partial \over \partial \gamma} 
{1 \over \sinh \eta}
\:.\end{eqnarray}
Summing over all the bonds of the graph and using the following identities:
\be
\sum_{(ab)}\left( (T_{aa}+T_{bb})\coth \eta_{ab}-2 
{T_{ba} \over \sinh \eta_{ab}} \right )= \tr{TM} = V
\:,\ee
\be
 \sum_{(ab)} \left( (T_{aa}+T_{bb}) {\partial \over \partial \gamma} \coth \eta_{ab} -2 T_{ba} 
{\partial \over \partial \gamma}{1 \over \sinh \eta_{ab}} \right)= \tr{T {\partial \over \partial \gamma} M} 
\:,\ee
\be
\eta \coth \eta =2 \gamma {\partial \over \partial \gamma} \ln \sinh \eta
\:,\ee
we find that the sum $P=\int_{\rm Graph}\D{y}\, G(y,y,)$ 
simplifies considerably into:
\be
P={\partial \over \partial \gamma} \sum_{(ab)} \ln \sinh \eta_{ab} + 
{V-B \over 2 \gamma} + \tr{M^{-1}{\partial \over \partial \gamma}M}
\:.\ee
Using the following property
$
\tr{M^{-1} {\partial \over \partial \gamma} M}=
{\partial \over \partial \gamma} \ln \det (M)
$
one finally obtains:
\be
P= {\partial \over \partial \gamma}\ln S(\gamma)
\ee
where
\be
S(\gamma)= \gamma^{{V-B \over 2}} 
\prod _{(ab)} \sinh \eta_{ab} \det(M)
\label{formule5}
\:.\ee


\section{The symmetry of the coefficients of the diagrammatic expansion
         \label{SymmCoef}}

The purpose of this appendix is to discuss the symmetry properties between
the coefficients appearing in the expansion of the determinant discussed in
section \ref{DiagExp}. 
This symmetry, a consequence of properties of the matrix 
$E$, considerably symplifies the calculation of the spectral 
determinant by the diagrammatic method. 
Since the presence of a magnetic field does not affect any point of the 
following discussion, one will forget it.
One will start to consider first the case of Neumann boundary conditions and
will indicate at the end of the appendix how to extend the result to the 
more general case.

Let us write $\det(1-E)=\det(E)\det(1-E^{-1})$. The inverse of $E$ is easily 
calculated since $\epsilon^{-1}=\epsilon^{\rm T}$ and $D$ is diagonal:
$D^{-1}_{ij}=\delta_{ij}\EXP{\sqrt\gamma l_i}$. Then one has 
\be\label{Symm}
\det(1-E)=\det(1-\epsilon D)
=(-1)^{V-B}\EXP{-2\sqrt\gamma L}\det(1-\epsilon D^{-1})
\ee
(for the calculation of $\det(E)$, see appendix \ref{MatProp}).
The matrix $\epsilon D^{-1}$ is equal to the matrix $E$ in which one has 
replaced $\sqrt\gamma$ by $-\sqrt\gamma$. This implies that
$\det(1-\epsilon D)$ and $\det(1-\epsilon D^{-1})$ exhibit similar 
expansions (with the same numerical factors).

To use this relation it is convenient to organize the expansion of 
$\det(1-E)$ in terms associated with diagrams or product of diagrams involving
the same set of arcs. Let us call $g_n=\{i_1,\cdots,i_n\}$ a set of $n$ arcs
of ${\cal G}$ (in particular $g_0=o\hspace{-0.175cm}/$ and $g_{2B}={\cal G}$). 
One may write
\be\label{expleng}
\det(1-E)=\sum_{n=0}^{2B}\sum_{g_n}\kappa[g_n]\EXP{-\sqrt\gamma\,l[g_n]}
\:,\ee
where $l[g_n]= l_{i_1}+\cdots+l_{i_n}$. The coefficient $\kappa[g_n]$ 
is a sum of coefficients $\alpha(\tilde{C})$ or a sum of product of such 
coefficients.
Equation (\ref{Symm}) implies that there is a relation between the terms 
involving diagrams with a large number of arcs and a small number of arcs.
More precisely, as a consequence of the obvious relation
$l[{\cal G}- g_n]=2L-l[g_n]$, we obtain
\be\label{symcoef}
\kappa[{\cal G}- g_n]=(-1)^{V-B}\kappa[g_n]
\:,\ee
where ${\cal G}- g_n=\{i/i\notin g_n\}$ is the set of all arcs of ${\cal G}$ 
except those of $g_n$.
As a simple consequence, the term of $\det(1-E)$ related to the diagrams of 
length $2L$ is precisely $(-1)^{V-B}\EXP{-2\sqrt\gamma L}$ since 
$\kappa[{\cal G}]=(-1)^{V-B}\kappa[o\hspace{-0.175cm}/]$

The use of the relation (\ref{symcoef}) is particularly powerful since it 
allows one to consider only half of the expansion (\ref{expleng}). 
Moreover the terms $\kappa[g_n]\EXP{-\sqrt\gamma\,l[g_n]}$ for small $n$ not 
only involve a small number of diagrams but also the simplest diagrams, 
constructed with less than $B$ arcs.
Taking into account in only the $B$ first terms in (\ref{expleng}) 
considerably reduce the number of diagrams to be considered.

As an example, one may apply this relation for the graph of figure 
\ref{ouvrebouteille}. For  the term $n=1$ in (\ref{expleng}), the sum over
$g_1$ brings four ensembles $\{1\}$, $\{\bar1\}$, $\{2\}$ and $\{\bar2\}$. 
Only the two first give some contributions since $\{2\}$ and $\{\bar2\}$ are 
not associated with any orbit.
Consider the case where $g_1=\{1\}$. Then the term 
$\kappa[\{1\}]\EXP{-\sqrt\gamma\,l[\{1\}]}$, given by the diagram
\be
\diagram{po1.eps}{0.75}{-0.3cm}
\:,\ee
has the same numerical factor as 
$\kappa[\{\bar1,2,\bar2\}]\EXP{-\sqrt\gamma\,l[\{\bar1,2,\bar2\}]}$
associated with the two following diagrams
\be
\diagram{po6.eps}{0.75}{-0.3cm}+\diagram{po23.eps}{0.75}{-0.3cm}
\:.\ee
In (\ref{expansion}) only the 3 first diagrams, among the 15, have to be 
considered and all other terms are given by the relation (\ref{symcoef}). 
It is also possible to check the symmetry (\ref{symcoef}) on equation 
(\ref{detob1}).

\bigskip

To end this appendix let us explain how the relation (\ref{symcoef}) is 
generalized for mixed boundary conditions. In this case the inverse of 
$\epsilon$ is given by $\epsilon^{\rm T}$ in which one has replaced
$\sqrt\gamma$ by $-\sqrt\gamma$ (the coefficients $\alpha(C)$ and
$\kappa[g_n]$ now depend on $\sqrt\gamma$).
One has 
$\det(1-E)=\det(E)\,\big[
\det(1-\epsilon D)\big|_{\sqrt\gamma \to -\sqrt\gamma}\big]$. 
Then equation (\ref{symcoef}) reads in the more general case
\be
\kappa[{\cal G}- g_n]=\det(\epsilon)\,
\kappa[g_n]\big|_{\sqrt\gamma \to -\sqrt\gamma}
\:,\ee
$\det(\epsilon)$ being given by (\ref{deteps}).


\section{Precisions on loops and multiple bonds between vertices\label{loops}}

\begin{figure}[!h]
\begin{center}
\includegraphics{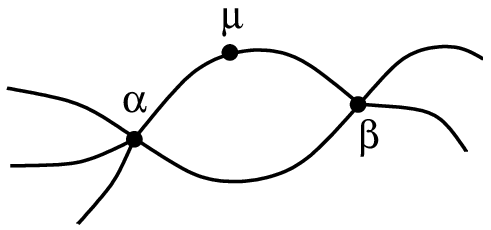}
\end{center}
\caption{}
\label{2bonds}
\end{figure}

\begin{figure}[!h]
\begin{center}
\includegraphics{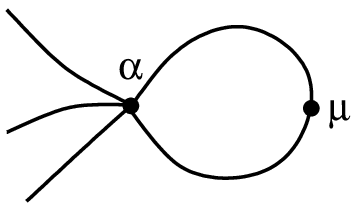}
\end{center}
\caption{}
\label{loop}
\end{figure}

In the section \ref{DefNot} we have assumed that a bond links always two 
different vertices and that there is only one bond between two different 
vertices. 
Formula (\ref{M1},\ref{M2}) apply to those kind of graph. As we remarked in 
a footnote, if two vertices are linked by two bonds, one can introduce a 
vertex in one of those bonds to fall back to the situation of section 
\ref{DefNot} (see figure \ref{2bonds}). Similarly if a bond starts and 
finishes at the same vertex to form a loop it is possible to separate it 
into two bonds by introducing an additional vertex (see figure \ref{loop}). 
Of course, these 
operations do not change the spectral properties of the graph. Any graph
can be described as in section \ref{DefNot} but the price to pay is to add
some vertices with coordinence $2$, which increases $V$ and $B$. In sections 
\ref{TraFor} and \ref{DiagExp} one has already noticed that vertices of
coordinence $2$ play no role in the construction of orbits. On the other hand,
since the sizes of the matrices that we have introduced are related to $B$ and
$V$, the calculations would become easier if it would be possible to minimize
those numbers. It is the purpose of this section to explain how to generalize
(\ref{M1},\ref{M2}) to the more general case where a loop can be present
at a vertex and several bonds link two vertices. A discussion is also given 
in \cite{Pas98} using a different method. We set $\gamma=1$.

\noindent
({\it i}) The expression (\ref{sp}) is again a good starting point. One 
considers the determinant of a graph one part of which is represented in 
figure \ref{2bonds}. In (\ref{sp}), the terms that involve vertex $\mu$ are:
\be
S(\gamma)^{-1}=\int\cdots\int\D\phi_\mu\D\bar\phi_\mu
G_{l_{\beta\mu}}(\vec\phi_\beta,\vec\phi_\mu)\,
G_{l_{\mu\alpha}}(\vec\phi_\mu,\vec\phi_\alpha)\cdots
\ee
Since $\phi_\mu$ does not appear anywhere else one can integrate over 
$\phi_\mu$, using the completeness relation for the propagator. One finds
\be\label{afint}
S(\gamma)^{-1}=\int\cdots
G_{l_{\beta\mu}+l_{\mu\alpha}}(\vec\phi_\beta,\vec\phi_\alpha)\cdots
\ee
In (\ref{afint}), the integral contains two propagators that propagate the 
field from vertex $\alpha$ to vertex $\beta$: 
$G_{l_{\beta\mu}+l_{\mu\alpha}}(\vec\phi_\beta,\vec\phi_\alpha)$ and
$G_{\lab}(\vec\phi_\beta,\vec\phi_\alpha)$.
It is now easy to see what is the natural generalization of 
(\ref{M1},\ref{M2}) if $B_{\alpha\beta}$ bonds of lengths $l_{\alpha\beta}^j$ 
link the two vertices:
\bea
M_{\alpha\alpha} &=& \sum_{j=1}^{B_{\alpha\beta}}
\coth(\sqrt\gamma l_{\alpha\beta}^j)+\cdots \\
M_{\alpha\beta}  &=& -\sum_{j=1}^{B_{\alpha\beta}}
\frac{\EXP{\I\theta_{\alpha\beta}^j}}{\sinh(\sqrt{\gamma}\lab^j)}
\:.\eea

Let us stress that the elimination of a vertex of coordinence $m_\mu=2$ is 
particularly straightforward with expression (\ref{sp}), since it is a 
consequence of the completeness relation for the propagtor of the 
two-dimensional harmonic oscillator.

\noindent
({\it ii}) If one now considers a graph as in figure \ref{loop}, 
one can use the same trick as before. It is then easy to see that the 
contribution of the loop is 
\be
S(\gamma)^{-1}=\int\cdots
G_{l_{\alpha\mu}^1+l_{\alpha\mu}^2}
(\phi_\alpha\EXP{-\I\theta_{\alpha\mu}^1},
 \phi_\alpha\EXP{-\I\theta_{\alpha\mu}^2})\cdots
\ee
where $l_{\alpha\mu}^1$ and $l_{\alpha\mu}^2$ are the lengths of the two 
bonds. Writing $l_{\alpha\alpha}=l_{\alpha\mu}^1+l_{\alpha\mu}^2$ for the 
length of the loop and 
$\theta_{\alpha\alpha}=\theta_{\alpha\mu}^1+\theta_{\mu\alpha}^2$ for the 
flux that pierces it, it is easy to see that the loop gives only a 
contribution to $M_{\alpha\alpha}$:
\be\label{Mwithloop}
M_{\alpha\alpha} = 2 \coth(\sqrt\gamma l_{\alpha\alpha}) -
                   2 \frac{\cos\theta_{\alpha\alpha}}
                          {\sinh(\sqrt{\gamma}l_{\alpha\alpha})} + \cdots
\ee

To conclude, gathering the points ({\it i}) and ({\it ii}) together one gives 
the general expression for $M$:
\be
M_{\alpha\beta} = \delta_{\alpha\beta}
\left[
  \sum_\mu a_{\alpha\mu}
  \sum_{j=1}^{B_{\alpha\mu}}
  \coth(\sqrt\gamma l_{\alpha\mu}^j) 
  +2\sum_{j=1}^{L_{\alpha}}
  \left(
    \coth(\sqrt\gamma l_{\alpha\alpha}^j)
   -\frac{\cos(\theta_{\alpha\alpha}^j)}
         {\sinh(\sqrt{\gamma}l_{\alpha\alpha}^j)}
  \right)
\right]
-a_{\alpha\beta} \sum_{j=1}^{B_{\alpha\beta}}
\frac{\EXP{\I\theta_{\alpha\beta}^j}}{\sinh(\sqrt{\gamma}\lab^j)}
\ee
where $L_\alpha$ is the number of loops at vertex $\alpha$ (we did not change
\footnote{this implies that $\sum_\beta a_{\alpha\beta}$ is no more the 
coordinence of vertex $\alpha$. One has: 
$m_\alpha=\sum_\beta a_{\alpha\beta}B_{\alpha\beta} + 2L_\alpha$.}
the definition of $a_{\alpha\beta}$ which  
is still $0$ or $1$ depending the vertices are connected or not; 
$a_{\alpha\alpha}=0$).
Let us remark that $M_{\alpha\beta}$ may be expressed in the more condensed
form \cite{DouRam85,Pas98}
\be
M_{\alpha\beta} = \delta_{\alpha\beta}\bigg[ 
  \sum_{\stackrel{{\rm arcs}\:b\:{\rm starting}}{\rm from\:vertex\:\alpha}}
  \coth(\sqrt\gamma l_{b}) - 2 \sum_{j=1}^{L_{\alpha}} 
  \frac{\cos(\theta_{\alpha\alpha}^j)}{\sinh(\sqrt{\gamma}l_{\alpha\alpha}^j)}
\bigg]
-a_{\alpha\beta} 
\sum_{\stackrel{{\rm arcs}\:b\:{\rm linking}}
               {\rm vertices\:\alpha\:and\:\beta}}
\frac{\EXP{\I\theta_b}}{\sinh(\sqrt{\gamma}l_b)}
\ee
where in the first sum over all arcs starting from vertex $\alpha$, the case 
of a loop brings twice the same contribution, as in (\ref{Mwithloop}), since 
it is associated with two arcs starting from the vertex.

As an example let us consider the graph of figure \ref{ouvrebouteille}. The 
number of vertices can be reduced to $2$ and the spectral determinant is 
given by a determinant of a $2\times2$ matrix:
\be
S(\gamma)=\sinh l\sinh b
\left|
  \begin{array}{cc}
  \coth b + 2\coth l - 2\frac{\cos\theta}{\sinh l}  & -\frac{1}{\sinh b} \\
  -\frac{1}{\sinh b}                                & \coth b
  \end{array}
\right|
\ee
which gives (\ref{Sob}).


\section{Some properties of the matrices $Q$, $R$, $D$, $\epsilon$ and $E$
         \label{MatProp}}

In the more general case, in the presence of a magnetic field and with 
mixed boundary conditions, the four matrices that we have introduced in 
section  \ref{Jean} read
\be\label{mq}
Q_{(\alpha\beta)(\mu\nu)}=a_{\alpha\beta}a_{\mu\nu}
\delta_{\alpha\mu}
\bigg(
  \frac{2}{m_\alpha+\frac{\lambda_\alpha}{\sqrt\gamma}}-\delta_{\beta\nu}
\bigg)\:\EXP{\I\frac{\theta_{\alpha\beta}-\theta_{\alpha\nu}}{2}}
,\:\ee
\be\label{mr}
R_{(\alpha\beta)(\mu\nu)}=a_{\alpha\beta}a_{\mu\nu}
\delta_{\alpha\nu}\delta_{\beta\mu}\EXP{-\sqrt\gamma\lab}
\:,\ee
\be\label{me}
\epsilon_{(\alpha\beta)(\mu\nu)}=a_{\alpha\beta}a_{\mu\nu}
\delta_{\beta\mu}
\bigg(
  \frac{2}{m_\beta+\frac{\lambda_\beta}{\sqrt\gamma}}-\delta_{\alpha\nu}
\bigg)
\ee
and
\be\label{md}
D_{(\alpha\beta)(\mu\nu)}=a_{\alpha\beta}a_{\mu\nu}
\delta_{\alpha\mu}\delta_{\beta\nu}
\EXP{-\sqrt{\gamma}\,l_{\mu\nu}+\I\theta_{\mu\nu}}
\:.\ee
The inverses of those matrices are easily calculated. They are given by their
hermitic conjugates in which one replaces $\sqrt\gamma$ by $-\sqrt\gamma$:
$Q^{-1}=Q^\dagger\big|_{\sqrt\gamma\to-\sqrt\gamma}$, {\it etc}
(let us recall that the transposition is defined as 
$Q^{\rm T}_{(\alpha\beta)(\mu\nu)}= Q_{(\mu\nu)(\alpha\beta)}$).
For example
\be
\epsilon^{-1}_{(\alpha\beta)(\mu\nu)}=a_{\alpha\beta}a_{\mu\nu}
\delta_{\alpha\nu}
\bigg(
  \frac{2}{m_\alpha-\frac{\lambda_\alpha}{\sqrt\gamma}}-\delta_{\beta\mu}
\bigg)
\:.\ee
This shows that in the positive part of the spectrum, when $\gamma=-k^2$, the 
five matrices $Q$, $R$, $D$, $\epsilon$ and $E$ are unitary matrices.
Let us notice that for Neumann boundary conditions $\lambda_\alpha=0$, then 
$Q^{-1}=Q$ and $\epsilon^{-1}=\epsilon^{\rm T}$.

We now compute the determinants of those matrices. To start with, 
one considers the sub-matrices $Q_\alpha$ defined in (\ref{Q}). We write the 
matrix element 
$(Q_\alpha)_{\beta\nu}=
\left(\frac{2}{m_\alpha+\lambda_\alpha/\sqrt\gamma}-\delta_{\beta\nu}\right)
\EXP{\I\frac{\theta_{\alpha\beta}-\theta_{\alpha\nu}}{2}}$,
where it is understood that $\beta$ and $\nu$ belong to the set of
$m_\alpha$ neighbours of $\alpha$. 

To compute the determinant of the matrix, the first step is to eliminate
all the phases with a unitary transformation. Then one has to consider a
$n\times n$ matrix of the form $F_{\alpha\beta}=a-\delta_{\alpha\beta}$.
It is possible to compute the determinant of such a matrix using a recursive 
method, one finds $\det(F)=(-1)^n(1-n\,a)$. Coming back to $Q_\alpha$ one 
has
$
\det(Q_\alpha)=(-1)^{m_\alpha+1}\,
\frac{m_\alpha-\lambda_\alpha/\sqrt\gamma}
     {m_\alpha+\lambda_\alpha/\sqrt\gamma}
$.
It follows that 
\be
\det(Q)=(-1)^V\prod_{\alpha=1}^V
\frac{m_\alpha-\lambda_\alpha/\sqrt\gamma}
     {m_\alpha+\lambda_\alpha/\sqrt\gamma}
\:.\ee
If $\gamma=-k^2$, this is indeed a complex number of unit modulus.

Next one would like to compute $\det(R)$. The matrix $R$ which couples 
time-reversed arcs has $2B$ non vanishing elements. The 
determinant of $R$ is then given by the product of all elements of $R$, 
times the sign of the permutation 
${\cal P}(1,\bar1,\cdots,B,\bar{B})=(\bar1,1,\cdots,\bar{B},B)$ 
that exchanges each arc with its time-reversed; it follows that 
\be
\det(R)=(-1)^B\EXP{-2\sqrt\gamma L}
\:.\ee

Since $\epsilon=({\cal U}^\dagger Q(\gamma)R(\gamma=0){\cal U})^{\rm T}$, it 
is easy to deduce its determinant:
\be\label{deteps}
\det(\epsilon)=(-1)^{V-B}\prod_{\alpha=1}^V
\frac{m_\alpha-\lambda_\alpha/\sqrt\gamma}
     {m_\alpha+\lambda_\alpha/\sqrt\gamma}
\ee
which involves the topological invariant $V-B$.

The determinant of $E$ is given by 
$\det(E)=\det(\epsilon)\EXP{-2\sqrt\gamma L}$. For Neumann boundary conditions
$\det(E)=(-1)^{V-B}\EXP{-2\sqrt\gamma L}$.

\end{appendix}



\vspace{1cm}


\noindent
{\bf E-mail:} \\
eric@physics.technion.ac.il \\
comtet@ipno.in2p3.fr    \\
desbois@ipno.in2p3.fr   \\
gilles@lps.u-psud.fr    \\
texier@kalymnos.unige.ch



\begin{thebibliography}{99}

\bibitem{RudSch53}
K.~Rudenberg and C.~Scherr,
\newblock J. Chem. Phys. {\bf 21}, 1565 (1953).

\bibitem{Ale83}
S.~Alexander,
\newblock Phys. Rev. B {\bf 27}(3), 1541 (1983).

\bibitem{Ram84}
R.~Rammal,
\newblock J. Phys. I (France) {\bf 45}, 191 (1984).

\bibitem{AvrSad91}
J.~E. Avron and L.~Sadun,
\newblock Ann. Phys. (N.Y.) {\bf 206}, 440 (1991).

\bibitem{Avr95}
J.~E. Avron,
\newblock Adiabatic quantum transport,
\newblock in {\em Quantum Fluctuations}, edited by E.~Akkermans {\it et al}, 
  page 741, Proceedings of the Les Houches
  Summer School, Session LXI, 1995, Elsevier, Amsterdam.

\bibitem{DouRam85}
B.~Dou{\c c}ot and R.~Rammal,
\newblock Phys. Rev. Lett. {\bf 55}(10), 1148 (1985).

\bibitem{DouRam86}
B.~Dou{\c c}ot and R.~Rammal,
\newblock J. Physique {\bf 47}, 973--999 (1986).

\bibitem{Mon96b}
G.~Montambaux,
\newblock Spectral properties of disordered conductors,
\newblock in {\em Quantum Fluctuations}, edited by S.~Reynaud {\it et al}, 
  page 387, Proceedings of the Les Houches Summer School,
  Session LXIII, 1996, Elsevier, Amsterdam.

\bibitem{PasMon98}
M.~Pascaud and G.~Montambaux,
\newblock Phys. Uspekhi {\bf 41}, 182 (1998).

\bibitem{Pas98}
M.~Pascaud,
\newblock {\em Magn\'etisme orbital de conducteurs m\'esoscopiques
  d\'esordonn\'es et propi\'et\'es spectrales de fermions en interaction},
\newblock PhD thesis, Universit\'e Paris XI, 1998.

\bibitem{PasMon99}
M.~Pascaud and G.~Montambaux,
\newblock Phys. Rev. Lett. {\bf 82}, 4512 (1999).

\bibitem{KotSmi97}
T.~Kottos and U.~Smilansky,
\newblock Phys. Rev. Lett. {\bf 79}(24), 4794 (1997).

\bibitem{KotSmi99}
T.~Kottos and U.~Smilansky,
\newblock Ann. Phys. (N.Y.) {\bf 274}(1), 76 (1999).

\bibitem{KotSmi99a}
T.~Kottos and U.~Smilansky,
\newblock (1999),
\newblock preprint chao-dyn/9906008.

\bibitem{ItzDro89a}
C.~Itzykson and J.-M. Drouffe,
\newblock {\em Statistical field theory}, volume~2, chapter~11,
\newblock Cambridge University Press, 1989.

\bibitem{Big76}
N.~Biggs,
\newblock {\em Algebraic Graph Theory},
\newblock Cambridge University Press, 1976.

\bibitem{CveDooSac80}
D.~M. Cvetkovic, M.~Doob, and H.~Sachs,
\newblock {\em Spectra of graphs, Theory and Application},
\newblock Academic Press, 1980.

\bibitem{Chu97}
F.~Chung,
\newblock {\em Spectral Graph Theory},
\newblock American Mathematical Society, 1997.

\bibitem{Col98}
Y.~{Colin de Verdi\`ere},
\newblock {\em Spectres de Graphes},
\newblock Societe Mathematique de France, 1998.

\bibitem{Car99}
R.~Carlson,
\newblock Inverse eigenvalue problems on directed graphs,
\newblock Transaction of the American Mathematical Society  (1999).

\bibitem{Rot83}
J.-P. Roth,
\newblock C. R. Acad. Sc. Paris {\bf 296}, 793 (1983).

\bibitem{Rot83a}
J.-P. Roth,
\newblock in {\em Colloque de Th\'eorie du potentiel - Jacques Deny}, page 521,
  Orsay, 1983.

\bibitem{Sel56}
A.~Selberg,
\newblock J. Ind. Math. Soc. {\bf 20}, 47 (1956).

\bibitem{Gut90}
M.~Gutzwiller,
\newblock in {\em Chaos in Classical and Quantum Mechanics}, edited by F.~John,
  volume~1 of {\em Interdisciplinary Applied Mathematics}, Springer Verlag, New
  York, 1990.

\bibitem{Ber84}
G.~Bergmann,
\newblock Phys. Rep. {\bf 107}, 1 (1984).

\bibitem{AslLar74}
L.~G. Aslamasov and A.~I. Larkin,
\newblock Sov. Phys. JETP {\bf 40}, 321 (1974).

\bibitem{Mon96}
G.~Montambaux,
\newblock J. Phys. I (France) {\bf 6}, 1 (1996).

\bibitem{Eck91}
U.~Eckern,
\newblock Z. Phys. B {\bf 82}, 393 (1991).

\bibitem{UllRicBarOppJal97}
D.~Ullmo, K.~Richter, H.~Baranger, F.~{von~Oppen}, and R.~Jalabert,
\newblock Physica E {\bf 1}, 268 (1997).

\bibitem{ArgImrSmi93}
N.~Argaman, Y.~Imry, and U.~Smilansky,
\newblock Phys. Rev. B {\bf 47}, 4440 (1993).

\bibitem{FeyHib65}
R.~P. Feynman and A.~R. Hibbs,
\newblock {\em Quantum mechanics and path integrals},
\newblock McGraw-Hill, 1965.

\bibitem{Eck93}
B.~Eckhardt,
\newblock Periodic orbit theory,
\newblock in {\em Quantum Chaos}, edited by G.~Casati {\it et al}, 
  page~77, Proceedings of the Int. School of Physics "Enrico
  Fermi", Course CXIX, 1993, North Holland.

\bibitem{But85}
M.~B{\"u}ttiker,
\newblock Phys. Rev. B {\bf 32}(3), 1846 (1985).

\bibitem{AkkAueAvrSha91}
E.~Akkermans, A.~Auerbach, J.~E. Avron, and B.~Shapiro,
\newblock Phys. Rev. Lett. {\bf 66}(1), 76 (1991).

\end{thebibliography}
\end{document}